\documentclass[twocolumn]{autart}

\usepackage{amsfonts}
\usepackage{amsmath}
\usepackage{amssymb}
\usepackage{amsxtra}
\usepackage{mathtools}
\usepackage[agsmcite]{harvard}
\usepackage{cite}
\usepackage{bm}
\usepackage[usenames,dvipsnames]{color}
\usepackage{epsf}
\usepackage{epsfig}
\usepackage{epstopdf}
\usepackage{graphicx}
\usepackage{listings}
\usepackage{mathrsfs}
\usepackage{times}
\usepackage{url}
\usepackage{verbatim}
\usepackage{multicol}
\usepackage{subfigure}
\usepackage{enumerate}
\usepackage{float}
\usepackage{algorithm}
\usepackage{etoolbox}  
\usepackage{arydshln}
\usepackage{txfonts}  

\let\classAND\AND
\let\AND\relax
\usepackage{algorithmic}

\let\AND\classAND
\AtBeginEnvironment{algorithmic}{\let\AND\algoAND}

\newenvironment{proof}{\textbf{Proof.}}{\hspace{\stretch{1}}\rule{1ex}{1ex}}   
\newtheorem{lemma}{Lemma}                                                    
\newtheorem{theorem}{Theorem}                                                
\newtheorem{example}{Example}
\newtheorem{assumption}{Assumption}
\newtheorem{proposition}{Proposition}
\newtheorem{problem}{Problem}
\newtheorem{primal-pro}{Primal Problem}
\newtheorem{dual-pro}{Dual Problem}
\newtheorem{auxiliary-pro}{Auxiliary Problem}

\date{}

\newcommand{\tr}{\mathrm{Tr}}

\newcommand{\vech}{\mathrm{vech}}
\newcommand{\vecs}{\mathrm{vecs}}
\newcommand{\vecn}{\mathrm{vecn}}

\newcommand{\st}{\mathrm{s.t.}}
\graphicspath{{figures/}}
\renewcommand{\appendixname}{Appendix~\Alph{section}}

\begin{document}
\graphicspath{{figures/}}
\begin{frontmatter}
	
	\title{Model-Free Design of Stochastic LQR Controller from Reinforcement Learning and Primal-Dual Optimization Perspective} 
	\vspace{-0.3cm}
	
	\vspace{-0.5cm}
	\author[USTC]{Man Li}\ead{lm1994@mail.ustc.edu.cn},    
	\author[USTC]{Jiahu Qin}\ead{jhqin@ustc.edu.cn},               
	\author[WSU]{Wei Xing Zheng}\ead{w.zheng@westernsydney.edu.au},  
	\author[HNU,HNU1]{Yaonan Wang}\ead{yaonan@hnu.edu.cn},  
	\author[USTC]{Yu Kang}\ead{kangduyu@ustc.edu.cn}
	
	\vspace{-0.2cm}
	\address[USTC]{Department of Automation, University of Science and Technology of China, Hefei 230027, China}  
	\vspace{-0.2cm}
	\address[WSU]{School of Computing, Engineering and Mathematics, Western Sydney University, Sydney, NSW 2751, Australia}        
	\vspace{-0.2cm}
	\address[HNU]{College of Electrical and Information Engineering, Hunan University, Changsha 410082, China}    
	\vspace{-0.2cm} 
	\address[HNU1]{National Engineering Laboratory for Robot Visual Perception and Control Technology, Changsha 410082, China}   

	\vspace{-0.8cm}
	
	\begin{abstract}                          
		To further understand the underlying mechanism of various reinforcement learning (RL) algorithms and also to better use the optimization theory to make further progress in RL, many researchers begin to revisit the linear-quadratic regulator (LQR) problem, whose setting is simple and yet captures the characteristics of RL.	
		Inspired by this, this work is concerned with the model-free design of stochastic LQR controller for linear systems subject to Gaussian noises, from the perspective of both RL and primal-dual optimization.
		From the RL perspective, we first develop a new model-free off-policy policy iteration (MF-OPPI) algorithm, in which the sampled data is repeatedly used for updating the policy to alleviate the data-hungry problem to some extent.
		We then provide a rigorous analysis for algorithm convergence by showing that the involved iterations are equivalent to the iterations in the classical policy iteration (PI) algorithm.
		From the perspective of optimization, we first reformulate the stochastic LQR problem at hand as a constrained non-convex optimization problem, which is shown to have strong duality.
		Then, to solve this non-convex optimization problem, we propose a model-based primal-dual (MB-PD) algorithm based on the properties of the resulting Karush-Kuhn-Tucker (KKT) conditions.
		We also give a model-free implementation for the MB-PD algorithm by solving a transformed dual feasibility condition.
		More importantly, we show that the dual and primal update steps in the MB-PD algorithm can be interpreted as the policy evaluation and policy improvement steps in the PI algorithm, respectively.
		Finally, we provide one simulation example to show the performance of the proposed algorithms.\vspace{-4mm}
	\end{abstract}
	
\end{frontmatter}

\section{Introduction}
\vspace{-2mm}
Reinforcement learning (RL), a branch of machine learning, aims to learn policies to optimize a long-term decision-making process in a dynamic environment \cite{sutton1998reinforcement,bertsekas2019reinforcement}.
In general, there are two main characteristics, which are also readily considered as the difficulties of the RL setting \cite{sutton1998reinforcement}.
First, the value of a state is the total amount of reward/cost an agent expects to accumulate over the future, starting from that state.
That is, early decisions affect the outcomes both near and far in the future.
Second, the learner's knowledge about the environment is incomplete, e.g., the underlying mechanism governing the dynamics of the state-transition trajectories is unknown.
To overcome these difficulties, a lot of effective RL algorithms have been developed \cite{Arulkumaran2017deep,Recht2019atour,qin2020multiplayer}, and these algorithms have been successfully used in numerous fields, including smart grids \cite{li2019distributed}, multi-robot systems \cite{wang2020mobile}, intelligent transportation\cite{Haydari2020deep}, and networked control \cite{Leong2020deep}.
From these works, we know that in spite of the fact that RL cannot be regarded as a standard optimization problem due to the lack of an explicit objective function \cite{naik2019discounted}, it still has strong ties to the optimization formulation, since its goal is to optimize the long-term cumulative performance \cite{zhang2019text}.
 In view of this, existing fruitful optimization algorithms and theories might enable researchers to make further progress in RL, so that the integration of elementary RL methods with optimization frameworks is worthwhile to study.
Some efforts towards this direction have been made, see, for example, \cite{qin2014sparse,Vieillard2019on}.
\vspace{-4mm}

It is worth mentioning that following this line, numerous researchers begin to revisit the linear-quadratic regulator (LQR) problem \cite{schluter2020event,Fazel2018global,Malik2019derivative,li2020distributed,karl2019finite,Yang2019provably}, the task of which is to determine an optimal feedback controller minimizing a quadratic cost function for linear systems \cite{mitze2020adynamic}.
One important reason for the revisit of the LQR problem is that it is simple and yet captures the two characteristics of RL problems mentioned in the previous paragraph.
Such properties of the LQR problem provide convenience for understanding the behavior and the underlying mechanism of various RL algorithms. 
Another reason is that the traditional methods for addressing the LQR problem require to directly solve an algebraic Riccati equation (ARE), which depends upon exact information of the system matrices; however, the exact system matrices are usually unknown in real applications.
Although model identification provides a way to obtain the estimated system model,  the propagation and accumulation of modeling errors may cause undesirable system behaviors.
Therefore, it is meaningful to study the model-free design of the LQR controller with the consideration of the integration of RL algorithms with optimization theories.
Some relevant works can be found in \cite{Fazel2018global,Malik2019derivative,Yang2019provably,karl2019finite,li2020distributed}.\vspace{-4mm}

Specifically, in \cite{Fazel2018global} and \cite{Malik2019derivative}, the authors reformulate the LQR problem as a non-convex optimization problem, and merge ideas from optimal control theory, mathematical optimization, and sampled-based RL to solve it.
To seek the solutions to the resulting non-convex optimization problem, model-free stochastic policy gradient algorithms along with zeroth-order optimization techniques are developed with provable global convergence \cite{Fazel2018global} and explicit convergence rate \cite{Malik2019derivative}.
The work \cite{li2020distributed} considers distributed RL for decentralized LQR problem, and proposes a zeroth-order distributed policy optimization algorithm to learn the local optimal controllers in a distributed manner, by leveraging the ideas of policy gradient, zeroth-order optimization, as well as consensus algorithms.
Note that in the above mentioned works, the key to realizing model-free control is to use the zeroth-order optimization approach to estimate the policy gradient.
This approach often suffers from large variance, as it involves the randomness of an entire trajectory.
To address this difficulty, some different algorithms have been proposed \cite{karl2019finite,Yang2019provably}.
The work \cite{karl2019finite} develops a model-free approximate policy iteration algorithm based on Q-function, in which the way to guarantee a good regret is also discussed.
In \cite{Yang2019provably}, an on-policy actor-critic algorithm, which is an online alternating update algorithm for bilevel optimization, is developed with a non-asymptotic convergence analysis, and then it is extended to the off-policy case to alleviate the data-hungry problem.
\vspace{-4mm}

Despite the fact  that extensive research has been carried out on the LQR problem when considering the integration of RL frameworks with optimization techniques, these works, such as \cite{Fazel2018global,Malik2019derivative,Yang2019provably,karl2019finite,li2020distributed}, mainly focus on the theoretical understanding of the performance of common RL algorithms.
They do not provide any explanations for the proposed RL algorithms from the perspective of optimization or establish any relationship between RL and optimization-based methods.
Different from the above mentioned works, the authors in \cite{lee2018primal} reformulate the LQR problem as a new optimization problem, and further explain its duality results from a Q-learning viewpoint.
The duality results proposed in \cite{lee2018primal} are based on the results given in \cite{david2001stochastic}, where the duality for the stochastic LQR problem pertaining to linear systems with stochastic noise is investigated in terms of semidefinite programming (SDP).
These works motivate some questions.
Can we formulate the stochastic LQR problem as an optimization problem whose optimal points can be interpreted from the perspective of various RL algorithms? 
Can we explain the RL algorithms used to solve stochastic LQR problem from the perspective of optimization?
\vspace{-4mm}


To answer these questions, we investigate the model-free design of stochastic LQR controller from the perspective of both RL and optimization.
We first solve the stochastic LQR problem in a model-free way from the perspective of RL.
Then, we reconsider this problem by some reformulation and transformations, and solve it from the perspective of primal-dual optimization.
Besides that, we also provide a theorem to show that the algorithms developed from the above two perspectives are equivalent in the sense that there is a corresponding relation in the involved iterations.
The contributions of this work are summarized as follows:\vspace{-4mm}
\begin{enumerate}
	\item To solve the stochastic LQR problem, we develop a model-free off-policy policy iteration (MF-OPPI) algorithm and a model-free primal-dual (MB-PD) algorithm from the perspective of RL and primal-dual optimization, respectively.
	Both algorithms can be implemented in a model-free way, so that one can circumvent model identification, which may cause undesirable behaviors due to the propagation and accumulation of modeling errors.
	More importantly, we show that the dual and primal update steps in the MB-PD algorithm can be interpreted as the policy evaluation and policy improvement steps in the classical policy iteration (PI) algorithm, respectively.
	This result provides a novel primal-dual optimization perspective to understand the classical PI algorithm.
	\item From the perspective of RL, we develop a new MF-OPPI algorithm to address the stochastic LQR problem, in which the data generated by exerting the behavior policy on the real system is repeatedly used to evaluate the target policy so as to alleviate the data-hungry problem to some extent. 
	Compared with the off-policy actor-critic algorithm proposed in \cite{Yang2019provably}, the newly developed MF-OPPI algorithm is simple and easy to implement, since it does not require to use neural networks for function approximation.
	\item From the optimization perspective, we first reformulate the stochastic LQR problem as a constrained non-convex optimization problem, based on the works \cite{lee2018primal} and \cite{david2001stochastic}.
	However, due to the consideration of the influence of stochastic noise and the relation with RL algorithms, the proposed optimization formulation is different from the ones given in \cite{lee2018primal} and \cite{david2001stochastic}.
	Such differences bring about some challenges in the equivalent formulation of primal-dual problems as well as the analysis for the strong duality.
	To address the resulting constrained non-convex optimization problem, we develop a new MB-PD algorithm with a model-free implementation, by using the properties of the resulting Karush-Kuhn-Tucker (KKT) conditions.
	This algorithm circumvents the requirement of probing noise exerted on the system for persistently exciting, which is necessary for the newly proposed MF-OPPI algorithm and other RL-based algorithms proposed in \cite{jiang2012computational,kiumarsi2017h,qin2018optimal}, thereby avoiding undesirable oscillation. 	
\end{enumerate}

\vspace{-4mm}The rest of this paper is organized as follows.
We first formulate the problem of interest and provide some preliminaries in Section~\ref{pro-for}.
Subsequently, we investigate this problem from the perspective of RL and primal-dual optimization in Section~\ref{sec-RL} and Section~\ref{sec-opt}, respectively.
Then, in Section~\ref{alg-equiv}, we show that the algorithms developed from the above two perspectives are equivalent in the sense that there is a corresponding relation in the involved iterations.
Next, we provide a simulation example in Section~\ref{simu}. 
Some concluding remarks are finally given in Section~\ref{concl}.
\vspace{-4mm}

\textit{Notations:} 
We use $\textbf{I}_n$ and $\textbf{0}_n$ to denote an identity matrix and a null matrix in $\mathbb{R}^{n\times n}$, respectively.
Denote by $0_n$ a zero vector of $n$ dimensions.
A set of symmetric matrices of $n\times n$ dimensions is denoted by $\mathbb{S}^n$. $\mathbb{S}^n_{+}$ ($\mathbb{S}^n_{++}$) is a set of symmetric positive semidefinite (positive definite) matrices.
For a matrix $M$, we use $M_{ij}$ to denote the element in the $i$-th row and the $j$-th column.
$\tr(M)$ is the trace of a matrix $M$, and $\rho(M)$ denotes its spectral radius.
A symmetric positive (semi-)definite matrix $M$ is denoted by $M\succ0\:(\succeq0)$.
\vspace{-4mm}

For a symmetric matrix $M\in\mathbb{R}^{n\times n}$, $\vech(M)\in\mathbb{R}^{1\times \frac{n(n+1)}{2}}$ denotes a row vector consisting of $n$ diagonal entries and $n(n-1)/2$ non-diagonal terms $M_{ij}$, $i=1,\hdots,n$, $i<j\le n$.
Specifically, $\vech(M)=[M_{11}, \, \hdots, \, M_{nn},\, M_{12},\,  \hdots,\, M_{1n}, M_{23},\, \hdots,\, M_{2n},\hdots,\, M_{(n-1)n}]$.
For a symmetric matrix $M\in\mathbb{R}^{n\times n}$,
$\vecs(M)\in \mathbb{R}^{n(n+1)/2}$ denotes a column vector composed of $n$ diagonal entries of $M$ and $n(n-1)/2$ distinct sums $M_{ij}+M_{ji}$, i.e.,
$\vecs(M)=[M_{11},  \hdots,\! M_{nn}, \!2M_{12},  \hdots, \! 2M_{1n},\! 2M_{23}, \hdots,\! 2M_{2n}, \hdots,\!2M_{(n-1)n}]^{\prime}$.
For a non-square matrix $M\in\mathbb{R}^{n\times m}$, $\vecn(M)\in\mathbb{R}^{mn}$ consists of the stack of each column's elements.
That is, $\vecn(M)=[M_{11}, \hdots,M_{n1},M_{12},\hdots,M_{n2},\hdots,M_{1m},\hdots,M_{nm}]^{\prime}$.
\vspace{-2mm}

\section{Problem Formulation and Preliminaries}\label{pro-for}
\vspace{-2mm}
Consider the following discrete-time linear time-invariant system\vspace{-2mm}
\begin{align}\label{LTI-syst}
x_{k+1}=Ax_k+Bu_k+w_k,
\end{align}

\vspace{-6mm}\noindent where $k\in\mathbb{N}$, $x_k\in\mathbb{R}^n$ is the state vector of the system plant, $u_k\in\mathbb{R}^m$ is the control action, $w_k\in\mathbb{R}^n$ is the system noise generated according to the Gaussian distribution $\mathcal{N}(0_n,\sigma_w^2)$, and $A$ and $B$ are constant matrices of appropriate dimensions.\vspace{-4mm}

The initial state is denoted by $x_0=z$, which is assumed to be a random variable.
In this paper, we consider the state-feedback control law, that is, $u_k=Fx_k$, where $F$ is the feedback gain.
$x_k(F,z)$ denotes the solution to the dynamical system \eqref{LTI-syst} under the control input $u_k=Fx_k$  while starting from $x_0=z$.
It is assumed that the noise $w_k$ is independent of the initial state $z$ as well as the noise $w_{k^{\prime}}$ for all $k^{\prime}\ne k$.\vspace{-4mm}

Under the state-feedback control law $u_k=Fx_k$, the cost function is denoted by\vspace{-2mm}
\begin{align}\label{cost}
J(F,z)\triangleq \sum_{k=0}^{\infty} \gamma^k \mathbb{E}\left\{
\left[
\begin{array}{c}
x_k(F,z)\\
Fx_k(F,z)
\end{array}
\right]^{\prime} \Lambda
\left[
\begin{array}{c}
x_k(F,z)\\
Fx_k(F,z)
\end{array}
\right]
\right\}
\end{align}

\vspace{-7mm}\noindent where $\gamma\in(0,1)$ is the discount factor, and  $\Lambda\triangleq\left[\begin{array}{cc}
Q& 0\\
0& R
\end{array}\right]\succeq 0$.\vspace{-4mm}

Define $\mathcal{F}\triangleq\{F\in\mathbb{R}^{m\times n}: \rho(A+BF)\le 1 \}$, which is the set of stabilizing state-feedback gains for the matrix pair $(A,B)$.
Now, we are ready to formally give the problem of interest, which is called the stochastic LQR problem.\vspace{-3mm}
\begin{problem}[Stochastic LQR Problem]\label{LQG-problem}
	Suppose that the initial state $z$ satisfying $\tilde{z}\triangleq\mathbb{E}(zz^{\prime})\succ 0$.
	The stochastic LQR problem aims to find the optimal feedback gain $F^*$ minimizing the quadratic cost function \eqref{cost} with the system state evolving along the dynamical system \eqref{LTI-syst}.
	That is,\vspace{-2mm}
	\begin{align*}
	\left\{
	\begin{aligned}
	\min_{F\in\mathcal{F}} & \quad J(F,z)\\
	\text{s.t.} & \quad x_{k+1}=Ax_k+Bu_k+w_k.
	\end{aligned}
	\right.
	\end{align*}
\end{problem}\vspace{-5.5mm}

The optimal feedback gain $F^*$ is defined as $F^* \triangleq\arg \min_{F\in\mathcal{F}}J(F,z) $.
The corresponding optimal cost for a given initial state $z\in\mathbb{R}^n$ is denoted by $J^*(z)\triangleq J(F^*,z)$.\vspace{-4mm}

The followings are some assumptions that will be used in the remaining contents.\vspace{-3mm}

\begin{assumption}\label{controllability}
	Assume that \vspace{-4mm}
	\begin{enumerate}
		\item $(A,B)$ is not known;
		\item An initial stabilizing feedback gain $F_{\text{stab}}\in\mathcal{F}$ is known;
		\item $Q\succeq 0$ and $R\succ 0$;
		\item The matrix pair $(A,B)$ is stabilizable;
		\item The matrix pair $(A,C)$ is detectable, where $C$ is a  matrix such that $CC^{\prime}=Q$.
	\end{enumerate}
\end{assumption}\vspace{-3mm}

For an stabilizing feedback gain $F\in\mathcal{F}$, define the value function as\vspace{-2mm}
\begin{align}\label{value-function}
V(F,x_k)\triangleq\sum_{i=k}^{\infty} \gamma^{i-k} \mathbb{E} [x_k^{\prime}(Q+F^{\prime}RF) x_k].
\end{align}

\vspace{-7mm}\noindent The optimal value function is given by \vspace{-2mm}
\begin{align}
V(F^*,x_k)\triangleq\min_{F\in\mathcal{F}} V(F,x_k).
\end{align}

\vspace{-7mm} \noindent \begin{lemma}[{\color{black}\cite[Lemma~3]{xiong2020model}}]
	For any stabilizing state-feedback gain $F\in\mathcal{F}$, the stochastic LQR problem, i.e., Problem~\ref{LQG-problem}, is well-posed\footnote{The stochastic LQR problem is said to be well-posed if the optimal value function satisfies $-\infty\le V(F^*,x_k)\le +\infty$.} and the corresponding value function is \vspace{-3mm}
	\begin{align}\label{quadratic-V}
	V(F,x_k)=\mathbb{E}[x_k^{\prime}Xx_k]+\frac{\gamma}{1-\gamma}\tr (X \sigma_w^2),
	\end{align}
	
	\vspace{-8mm}\noindent where $X\in\mathbb{S}_{+}^n$ is the unique solution to the Lyapunov equation \vspace{-3mm}
	\begin{align}\label{Lya-X}
	X=\gamma(A+BF)^{\prime}X(A+BF)+F^{\prime}RF+Q.
	\end{align}
\end{lemma}

\vspace{-7.2mm} From the definition of the value function \eqref{value-function}, one has \vspace{-3mm}
\begin{align}\label{Bellman-eq}
V(F,x_k)=\mathbb{E} [x_k^{\prime}(Q+F^{\prime}RF) x_k]+\gamma V(F,x_{k+1}),
\end{align}

\vspace{-7mm}\noindent which is called the Bellman equation.
Define the Hamiltonian as \vspace{-3mm}
\begin{align*}
H(F,x_k)\triangleq&\mathbb{E}[x_k^{\prime}(Q+F^{\prime}RF)x_k]+\gamma \mathbb{E}(x_{k+1}^{\prime}Xx_{k+1})\\
&-\mathbb{E}(x_k^{\prime}Xx_k)-\gamma \tr(X\sigma_w^2).
\end{align*}

\vspace{-7mm}\noindent The first-order necessary condition for optimality gives\vspace{-3mm}
\begin{align*}
\frac{\partial H(F,x_k)}{\partial F}=2(R+\gamma B^{\prime}XB)F\mathbb{E}[x_kx_k^{\prime}]+2\gamma B^{\prime}XA\mathbb{E}[x_kx_k^{\prime}]=0.
\end{align*}

\vspace{-7mm}\noindent It follows that the unique optimal control gain is given by \vspace{-3mm}
\begin{align}\label{optimal-F}
F^*=-\gamma(R+\gamma B^{\prime}X^*B)^{-1}B^{\prime}X^*A
\end{align}

\vspace{-7mm}\noindent with $X^*$ being the unique solution to the following ARE \vspace{-3mm}
\begin{align}\label{ARE}
X^*=Q+\gamma A^{\prime}X^*A-\gamma^2 A^{\prime}X^* B(R+\gamma B^{\prime}X^*B)^{-1}B^{\prime}X^*A.
\end{align}

\vspace{-7mm} \noindent By substituting the optimal feedback gain \eqref{optimal-F} into the value function \eqref{Bellman-eq}, one can obtain that \vspace{-3mm}
	\begin{align}\label{on-1}
	V(F^*,x_k)=\mathbb{E} \left\{\left[\begin{array}{c}
	x_k\\
	F^*x_k
	\end{array}
	\right]^{\prime}P^*\left[\begin{array}{c}
	x_k\\
	F^*x_k
	\end{array}
	\right]\right\}+\frac{\gamma}{1-\gamma} \tr(X^*\sigma_w^2),
	\end{align}
	
	\vspace{-10mm}\noindent where\vspace{-3mm}
	\begin{align}\label{P}
	P^*\triangleq \left[
	\begin{array}{cc}
	Q+\gamma A^{\prime}X^*A & \gamma A^{\prime}X^*B\\
	\gamma B^{\prime}X^*A  & R+\gamma B^{\prime}X^*B
	\end{array}
	\right].
	\end{align}
	
	\vspace{-7.2mm}\noindent In the following analysis, let $P_{11}^*\triangleq Q+\gamma A^{\prime}X^*A$, $P_{12}^*\triangleq(P_{21}^*)^{\prime}\triangleq\gamma A^{\prime}X^*B$, and $P_{22}^*\triangleq R+\gamma B^{\prime}X^*B$.\vspace{-3.8mm}

Based on the above analysis, we know that calculating the optimal feedback gain $F^*$ requires both the system matrix $A$ and the control input matrix $B$.
In the following two sections, we provide two approaches from a RL-based view and an optimization-based view, respectively, to compute the optimal feedback gain without using the matrices $A$ and $B$.\vspace{-2.5mm}

\section{RL-Based Method}\label{sec-RL}
\vspace{-2mm}
The goal of this section is to solve Problem~\ref{LQG-problem} by designing RL algorithms.\vspace{-2.5mm}
\subsection{Classical PI Algorithm}
\vspace{-2mm}

\begin{algorithm}
	\caption{Classical PI Algorithm}
	\label{alg-offline}
	\begin{algorithmic}[1]
		\STATE \textbf{Initialization:} set a stabilizing control gain $F^{0}$, $s=0$, and $\epsilon>0$;
		\STATE \textbf{Repeat}
		\STATE \quad  \textbf{policy evaluation:} solve $X^s$ from \vspace{-2mm}
				\begin{equation}
				\begin{aligned}\label{-PE}
				 \  X^s=\gamma(A+BF^s)^{\prime}X^s(A+BF^s)+(F^s)^{\prime}RF^s+Q;
				\end{aligned}
				\end{equation}	
		\STATE \vspace{-2mm} \quad \textbf{policy improvement:} update control gain from \vspace{-2mm}
				\begin{align}\label{-PI}
				 F^{s+1}=-\gamma(R+\gamma B^{\prime}X^sB)^{-1}B^{\prime}X^sA;
				\end{align}		
		\STATE \vspace{-2mm} \quad $s\leftarrow s+1;$
		\STATE \textbf{Until} $\|F^{s}-F^{s-1}\|\le\epsilon$;
		\STATE \textbf{Return} $F^s$.
	\end{algorithmic}
\end{algorithm}

Note that by replacing $(A,B)$ with $(\gamma^{1/2}A,\gamma^{1/2}B)$, the ARE \eqref{ARE} is identical to that obtained from a standard LQR problem \cite[Chapter~2.1.4]{vrabie2013optimal}, in which the dynamical system is not influenced by the stochastic noise.
There have been various algorithms to seek the solutions to the ARE, one of which is the policy iteration algorithm.
A classical PI algorithm proposed in \cite[Chapter~2.2]{vrabie2013optimal} to solve \eqref{ARE} is given below, cf. Algorithm~\ref{alg-offline}. \vspace{-4mm}

%
%

The iterations \eqref{-PE} and \eqref{-PI} in Algorithm~\ref{alg-offline} rely upon repeatedly solving the Lyapunov equation \eqref{Lya-X}. 
Hence, Algorithm~\ref{alg-offline} is an offline algorithm that requires complete knowledge of the system matrices $A$ and $B$.
Note that Algorithm~\ref{alg-offline} is also called Hewer's algorithm, and it has been shown to converge to the solution of the ARE \eqref{ARE} in \cite{hewer1971an}.
\vspace{-2mm}

\subsection{MF-OPPI Algorithm}
\vspace{-2mm}
In this subsection, we intend to propose an off-policy algorithm that can calculate the optimal feedback gain \eqref{optimal-F} without using the system matrices.\vspace{-4mm}

To this end, we first rewrite the original system \eqref{LTI-syst} as \vspace{-2mm}
\begin{align}\label{off-1}
x_{k+1}=(A+BF^s)x_k+B(u_k-F^sx_k)+w_k,
\end{align}

\vspace{-6mm} \noindent where $u_k^s\triangleq F^sx_k$ is the target policy being learned and updated by iterations, and $u_k$ is the behavior policy that is exerted on the system to generate data for learning.\vspace{-4mm}

By using \eqref{off-1} and the Lyapunov equation \eqref{Lya-X}, one can obtain that \vspace{-2mm}
\begin{equation}
\begin{aligned}\label{off-2}
&\mathbb{E}[x_k^{\prime}X^sx_k]-\gamma \mathbb{E}[x_{k+1}^{\prime}X^sx_{k+1}]\\
=&\mathbb{E}[x_k^{\prime}(Q+(F^s)^{\prime}RF^s)x_k]-2\gamma \mathbb{E}[(u_k-F^sx_k)^{\prime}B^{\prime}X^sAx_k]\\
&-\gamma \mathbb{E} [(u_k-F^sx_k)^{\prime}B^{\prime}X^sB(F^sx_k+u_k)]-\gamma \tr(X^s \sigma_w^2),
\end{aligned}
\end{equation}

\vspace{-5mm}\noindent which is termed the off-policy Bellman equation.\vspace{-4mm}

Then, we use \eqref{off-2} for policy evaluation, and propose an MF-OPPI algorithm, i.e., Algorithm~\ref{alg-off-policy}.
\begin{algorithm}
	   \caption{MF-OPPI Algorithm}
		\label{alg-off-policy}
	\begin{algorithmic}[1]
		\STATE \textbf{Initialization:} set a stabilizing control gain $F^{0}$, $s=0$, and $\epsilon>0$;
		\STATE \textbf{Repeat}
		\STATE \quad  \textbf{policy evaluation:} solve $X^s$, $X_1^s$, and $X_2^s$ from \vspace{-3mm}
		\begin{equation}
		\begin{aligned}\label{off-PE}
		 &\mathbb{E}[x_k^{\prime}X^sx_k]-\gamma \mathbb{E}[x_{k+1}^{\prime}X^sx_{k+1}]\\
		=&\mathbb{E}[x_k^{\prime}(Q+(F^s)^{\prime}RF^s)x_k]-2\gamma \mathbb{E}[(u_k-F^sx_k)^{\prime}X_1^sx_k]\\
		&-\gamma \mathbb{E} [(u_k-F^sx_k)^{\prime}X_2^s(F^sx_k+u_k)]-\gamma  \tr(X^s \sigma_w^2);
		\end{aligned}
		\end{equation}		
		\STATE \vspace{-0mm}\quad \textbf{policy improvement:} update control gain from\vspace{-3mm}
		\begin{align}\label{off-PI}
		F^{s+1}=-\gamma(R+\gamma X_2^s)^{-1}X_1^s;
		\end{align}
		\STATE \vspace{-3mm} \quad $s\leftarrow s+1;$
		\STATE \textbf{Until} $\|F^{s}-F^{s-1}\|\le\epsilon$;
		\STATE \textbf{Return} $F^s$.
	\end{algorithmic}
\end{algorithm}\vspace{-2mm}

%
%
%

\begin{theorem}\label{thm-off}
	In Algorithm~\ref{alg-off-policy}, it holds that $\lim_{s\to\infty}X^s=X^*$ and $\lim_{t\to\infty}F^s=F^*$, where $X^*$ and $F^*$ are defined in \eqref{ARE} and \eqref{optimal-F}, respectively.
\end{theorem}\vspace{-2mm}
\begin{proof}
	Substituting \eqref{off-1} into the left-hand side of \eqref{off-PE}, it follows that\vspace{-2mm}
	\begin{align*}
	&\mathbb{E}[x_k^{\prime}X^sx_k]-\gamma \mathbb{E}[x_{k+1}^{\prime}X^sx_{k+1}]\\
	=&\mathbb{E}[x_k^{\prime}X^sx_k]-\gamma \mathbb{E}[((A+BF^s)x_k+w_k)^{\prime}X^s((A+BF^s)x_k+w_k)]\\
	&-2\gamma \mathbb{E}[(u_k-F^sx_k)^{\prime}B^{\prime}X^sAx_k]\\
	&-\gamma \mathbb{E} [(u_k-F^sx_k)^{\prime}B^{\prime}X^sB(F^sx_k+u_k)].
	\end{align*}
	
	\vspace{-6mm}\noindent Therefore, it holds that \vspace{-2mm}
	\begin{align*}
	&\mathbb{E}[x_k^{\prime}X^sx_k]-\gamma \mathbb{E}[((A+BF^s)x_k+w_k)^{\prime}X^s((A+BF^s)x_k+w_k)]\\
	=&\mathbb{E}[x_k^{\prime}(Q+(F^s)^{\prime}RF^s)x_k]-\gamma \tr(X^s \sigma_w^2).
	\end{align*}
	
	\vspace{-6mm}\noindent Since $\mathbb{E}[((A+BF^s)x_k+w_k)^{\prime}X^s((A+BF^s)x_k+w_k)]=\mathbb{E}[x_k^{\prime}(A+BF^s)^{\prime}X^s(A+BF^s)x_k]+\tr(X^s \sigma_w^2)$, one can obtain that for $k\ge 1$, \vspace{-2mm}
	\begin{equation}
	\begin{aligned}\label{off-3}
	&\mathbb{E}[x_k^{\prime}X^sx_k]-\gamma \mathbb{E}[x_k^{\prime}(A+BF^s)^{\prime}X^s(A+BF^s)x_k]\\
	=&\mathbb{E}[x_k^{\prime}(Q+(F^s)^{\prime}RF^s)x_k].
	\end{aligned}
	\end{equation}
	
	\vspace{-6mm}\noindent From the dynamics \eqref{LTI-syst}, it holds that for $k\ge 1$, \vspace{-2mm}
	\begin{align*}
	\mathbb{E}[x_kx_k^{\prime}]=(A+BF)\mathbb{E}[x_{k-1}x_{k-1}^{\prime}](A+BF)^{\prime}+\sigma_w^2.
	\end{align*}
	
	\vspace{-6mm}\noindent Due to the positive definiteness of $\sigma_w^2$ and $\tilde{z}$, the above equation implies that $\mathbb{E}[x_kx_k^{\prime}]$ is positive definite.
	Thus, one can obtain from \eqref{off-3} that \vspace{-2mm}
	\begin{align}\label{off-8}
		X^s=\gamma(A+BF^s)^{\prime}X^s(A+BF^s)+(F^s)^{\prime}RF^s+Q, 
	\end{align}
	
    \vspace{-6mm}\noindent which is exactly the update in policy evaluation step of Algorithm~\ref{alg-offline}.  \vspace{-4mm}
	
	From \eqref{off-2} and \eqref{off-PE}, we have $X_1^s=B^{\prime}X^sA$ and $X_2^s=B^{\prime}X^sB$, which further imply that \eqref{off-PI} is identical to \eqref{-PI}. \vspace{-2mm}
\end{proof}

\subsection{Implementation of Algorithm~\ref{alg-off-policy}}
\vspace{-2mm}
To implement Algorithm~\ref{alg-off-policy}, a natural question is how to solve the matrices $X^s$, $X_1^s$, and $X_2^s$ from \eqref{off-PE}.
In this subsection, we provide an effective way for the implementation of Algorithm~\ref{alg-off-policy}, where a numerical average is adopted to approximate the mathematical expectation and a batch least square (BLS) method \cite{graupe1980acom} is employed to estimate the unknown matrices. \vspace{-4mm}

Specifically, by vectorization, the policy evaluation step \eqref{off-PE} in Algorithm~\ref{alg-off-policy} becomes \vspace{-2mm}
\begin{align}\label{off-4}
&\left\{\mathbb{E} \left[\vech \big(x_kx_k^{\prime} \big)-\gamma \vech \big( x_{k+1}x_{k+1}^{\prime}  \big)  \right]+\gamma \vech( \sigma_w^2)\right\}\vecs(X^s) \notag\\
&+2\gamma \mathbb{E}\left[\vecn\big[(u_k-F^sx_k)x_k^{\prime}\big]\right]^{\prime}\vecn(X_1^s)\notag\\
&+\gamma \mathbb{E}\Big[\vech \big[(u_k-F^sx_k)(u_k+F^sx_k)^{\prime}\big]\Big]\vecs(X_2^s)\notag\\
=&\mathbb{E}\left[x_k^{\prime}(Q+(F^s)^{\prime}RF^s)x_k\right].
\end{align}

\vspace{-6mm}The equation \eqref{off-4} has $\frac{n(n+1)}{2}+mn+\frac{m(m+1)}{2}$ unknown parameters.
Thus, at least $K\ge\frac{n(n+1)}{2}+mn+\frac{m(m+1)}{2}$ sample points are required to solve $X^s$, $X_1^s$, and $X_2^s$ from \eqref{off-4}.
Sample the data generated under the behavior policy $u_k$ for $K$ time steps, and define \vspace{-2mm}
\begin{align}
\Phi^s\triangleq&\left[
\begin{array}{c}
\mathbb{E}[x_k^{\prime}(Q+(F^s)^{\prime}RF^s)x_k]\\
\mathbb{E}[x_{k+1}^{\prime}(Q+(F^s)^{\prime}RF^s)x_{k+1}]\\
\vdots \\
\mathbb{E}[x_{k+K-1}^{\prime}(Q+(F^s)^{\prime}RF^s)x_{k+K-1}]
\end{array}
\right],\label{off-5}\\
\Psi^s\triangleq&\left[
\begin{array}{ccc}
\mathbb{E}[H_{xx,1}] & \mathbb{E}[H_{xu,1}] & \mathbb{E}[H_{uu,1}]\\
\vdots   & \vdots   & \vdots  \\
\mathbb{E}[H_{xx,K}] & \mathbb{E}[H_{xu,K}] & \mathbb{E}[H_{uu,K}]
\end{array}
\right],\label{off-6}
\end{align}

\vspace{-6mm}\noindent with \vspace{-2mm}
\begin{align*}
H_{xx,i}\triangleq&\vech \big(x_{k+i-1}x_{k+i-1}^{\prime}  \big)-\gamma \vech \big( x_{k+i} x_{k+i}^{\prime}  \big) +\gamma \vech( \sigma_w^2),\\
H_{xu,i}\triangleq&2\gamma \vecn\left[(u_{k+i-1}-F^sx_{k+i-1})x_{k+i-1}^{\prime}\right]^{\prime},\\
H_{uu,i}\triangleq&\gamma \vech \ \Big  [(u_{k+i-1}-F^sx_{k+i-1})(u_{k+i-1}+F^sx_{k+i-1})^{\prime}\Big].
\end{align*}

\vspace{-6mm} From \eqref{off-4}, \eqref{off-5}, and \eqref{off-6}, one can obtain \vspace{-2mm}
\begin{align*}
\Psi^s\left[
\begin{array}{ccc}
\Big(\vecs(X^s)\Big)^{\prime} & \left(\vecn(X_1^s)\right)^{\prime} & \Big(\vecs(X_2^s)\Big)^{\prime}
\end{array}
\right]^{\prime}=\Phi^s,
\end{align*}

\vspace{-6mm}\noindent which can be solved by \vspace{-2mm}
\begin{equation}\label{off-7}
\begin{aligned}
&\left[
\begin{array}{ccc}
\Big(\vecs(X^s)\Big)^{\prime} &\Big(\vecn(X_1^s)\Big)^{\prime} &\Big(\vecs(X_2^s)\Big)^{\prime}
\end{array}
\right]^{\prime}\\
=&\left[(\Psi^s)^{\prime} \Psi^s\right]^{-1}(\Psi^s)^{\prime} \Phi^s.
\end{aligned}
\end{equation}

\vspace{-7mm} Note that to guarantee the solvability of the BLS estimator \eqref{off-7}, a probing noise is required to add into the behavior policy $u_k$ to guarantee the invertibility of  $(\Psi^s)^{\prime} \Psi^s$.
It is worthwhile to point out that adding a probing noise does not influence the iteration results of Algorithm~\ref{alg-off-policy}, which can be regarded as an advantage of the off-policy algorithm compared with the on-policy one \cite{kiumarsi2017h}.
The detailed implementation of Algorithm~\ref{alg-off-policy} is given in Algorithm~\ref{alg-imp-off-policy}. \vspace{-4mm}

Algorithm~\ref{alg-imp-off-policy} includes two main phases.
In the first phase, we exert the behavior policy on the real system for $K$ time steps, and sample the resulting system states and control actions.
Then, a numerical average of $N$ trajectories is used to calculate the mathematical expectation.
In the second phase, we implement policy evaluation and policy improvement steps by reusing the data sampled in the first phase.
The reuse of data is an important advantage of the proposed MF-OPPI algorithm when compared with the on-policy RL algorithms, since the latter are required to collect the state-input data when applying every iterated control policies \cite{qin2020multiplayer,vrabie2009adaptive}. Therefore, the proposed MF-OPPI algorithm is able to alleviate the data-hungry problem to some extent.

\begin{algorithm}
	\caption{Implementation of MF-OPPI Algorithm}  
	\label{alg-imp-off-policy}
	\begin{algorithmic}[1]
		\STATE \textbf{Phase~1 (Data Collection):} 
		\STATE \textbf{Initialization:} set a stabilizing control gain $F^{0}$, $s=0$, $K>0$, and $\epsilon>0$; let $E_{x_kx_k}=0$, $E_{x_ku_k}=0$, and $E_{u_ku_k}=0$ for $k=0,\cdots,K$;
		\STATE \textbf{For} $q=1:N$\\
		\STATE 	\quad apply the behavior policy $u_k$  for $K$ time steps;
		\STATE	\quad sample $\{x_k,u_k\}_{k=0,\hdots,K}$;
		\STATE	\quad $E_{x_kx_k}=E_{x_kx_k}+x_kx_k^{\prime}$, $k=0,\hdots,K$;
		\STATE  \quad $E_{x_ku_k}=E_{x_ku_k}+x_ku_k^{\prime}$, $k=0,\hdots,K$;
		\STATE	\quad $E_{u_ku_k}=E_{u_ku_k}+u_ku_k^{\prime}$, $k=0,\hdots,K$;
		\STATE \textbf{End}\\
	    \STATE	$E_{x_kx_k}=E_{x_kx_k}/N$, $E_{x_ku_k}=E_{x_ku_k}/N$, $E_{u_ku_k}=E_{u_ku_k}/N$, $k=0,\hdots,K$;
		\STATE \textbf{Phase~2 (Reuse of Collected Data for Iteration):}\\
		\STATE \textbf{Initialization:} set a stabilizable control gain $F^{0}$, $s=0$, and $\epsilon>0$;  
		\STATE \textbf{Repeat}	
		\STATE \quad calculate $\Phi^s$ and $\Psi^s$ using $F^s$ and collected data;\\	
		\STATE \quad \textbf{policy evaluation:} solve $X^s$, $X_1^s$, and $X_2^s$ from \vspace{-3mm}
			\begin{align*}
			 &\left[
			\begin{array}{ccc}
			\Big(\vecs(X^s)\Big)^{\prime} & \Big(\vecn(X_1^s)\Big)^{\prime} & \Big(\vecs(X_2^s)\Big)^{\prime}
			\end{array}
			\right]^{\prime}\\
			=&\left[(\Psi^s)^{\prime} \Psi^s\right]^{-1}(\Psi^s)^{\prime} \Phi^s;
			\end{align*}
		\STATE	\vspace{-3mm} \quad \textbf{policy improvement:} update control gain from\vspace{-2mm}
			$$F^{s+1}=-\gamma(R+\gamma B^{\prime}X^sB)^{-1}B^{\prime}X^sA;$$		
		\STATE	\vspace{-3mm} $s\leftarrow s+1$;
		\STATE \textbf{Until} $\|F^{s}-F^{s-1}\|\le\epsilon$
		\STATE \textbf{Return} $F^s$.
	\end{algorithmic}
\end{algorithm}

\section{Optimization-Based Method}\label{sec-opt}
\vspace{-2mm}
In this section, we reformulate Problem~\ref{LQG-problem} as a constrained optimization problem, and resolve it  from the primal-dual optimization perspective.

\subsection{Problem Reformulation}
\vspace{-2mm}
For technical reasons, we need to make some modifications on Problem~\ref{LQG-problem}.
To do this, some new notations are required.\vspace{-4mm}

We introduce the augmented state vector $v_k\triangleq\left[x_k^{\prime} \;\; u_k^{\prime} \right]^{\prime}$, and the resulting augmented system is described by \vspace{-2mm}
\begin{align}\label{augmented-system}
v_{k+1}=A_F v_k+\bar{F} w_k,
\end{align}

\vspace{-6mm}\noindent where $A_F\triangleq\left[
\begin{array}{cc}
A  &B\\
FA &FB
\end{array}\right]\in\mathbb{R}^{(n+m)\times(n+m)}$ and $\bar{F}\triangleq\left[
\begin{array}{c}
I_n\\
F
\end{array}\right]\in\mathbb{R}^{(n+m)\times n}$.\vspace{-4mm}

The following is a useful property that will be used in the remaining part of this paper.\vspace{-3mm}

\begin{lemma}[Property of Spectral Radius{\color{black}\cite[Lemma~1]{lee2018primal}}]\label{spectral}
	It holds that $\rho(A+BF)=\rho(A_F)$.
\end{lemma}\vspace{-3mm}

Let  $z_{(i)}\in\mathbb{R}^n$ for $i\in\{1,\hdots,r\}$ be $r$ different samples from the random variable $z$.
Let  $u_{(i)}\in\mathbb{R}^{m}$ for $i\in\{1,\hdots,r\}$ be $r$ different initial stabilizing control laws.
$r$ different initial states for the augmented system \eqref{augmented-system} are denoted by $v_{(i)}=[z_{(i)}^{\prime} \ u_{(i)}^{\prime}]^{\prime}$ for $i=\{1,\hdots,r\}$.
Denote by $v_k(F,v_{(i)})$ the solution of \eqref{augmented-system} starting from the initial state $v_{(i)}$ and evolving under the control $u_k=Fx_k$, $k\ge 1$.\vspace{-4mm}


A new cost function is defined as\vspace{-2mm}
\begin{align}\label{J_hat}
\hat{J}(F,v_{(i)})\triangleq \sum_{k=0}^{\infty}\gamma^k \mathbb{E}\left[ v_k(F,v_{(i)})^{\prime}\Lambda v_k(F,v_{(i)})\right].
\end{align}

\vspace{-7mm}\noindent Note that different from the cost function $J(F,z)$ defined in \eqref{cost}, where the initial state-feedback control law $u_0=Fz$ is considered, the initial control $u_{(i)}$ used in $\hat{J}(F,v_{(i)})$ is arbitrary, as long as it is stabilizable.

\vspace{-4mm} The new problem is formally formulated as below.\vspace{-3mm}
\begin{problem}[Modified Stochastic LQR Problem]\label{LQG-problem2}
	Suppose that $z_{(i)}\in\mathbb{R}^n$, $u_{(i)}\in\mathbb{R}^{m}$,  and $v_{(i)}=[z_{(i)}^{\prime} \ u_{(i)}^{\prime}]^{\prime}$ for $i\in\{1,\hdots,r\}$, are chosen such that  $\frac{1}{r}\sum_{i=1}^{r}v_{(i)}v_{(i)}^{\prime}=\Gamma\succ 0$, with $r$ being a positive constant.
	The following problem,\vspace{-2mm}
	\begin{align*}
	\left\{
	\begin{aligned}
	\min_{F\in\mathcal{F}} & \quad \frac{1}{r}\sum_{i=1}^{r}\hat{J}(F,v_{(i)})\\
	\text{s.t.} & \quad x_{k+1}=Ax_k+Bu_k+w_k,
	\end{aligned}
	\right.
	\end{align*}
	
	\vspace{-6mm}\noindent is called a modified stochastic LQR problem.
	The corresponding optimal feedback gain is defined as $\hat{F}^*\triangleq\arg\min_{F\in\mathcal{F}} \frac{1}{r}\sum_{i=1}^{r} \hat{J}(F,v_{(i)})$.
	The resulting optimal value is denoted by $\hat{J}^*\triangleq\frac{1}{r}\sum_{i=1}^{r} \hat{J}(F^*,v_{(i)})$.
\end{problem}\vspace{-3mm}

\begin{proposition}\label{equiv-F}
	The optimal solution $\hat{F}^*$ obtained by solving Problem~\ref{LQG-problem2} is unique, and it is identical to $F^*$.
\end{proposition}\vspace{-2mm}
\begin{proof}
	From the results in Section~\ref{pro-for}, we know that although $J^*(F,z)$ has different values under different initial state $z$, the optimal control gain $F^*=\arg\min_{F\in\mathcal{F}} J(F,z)$, whose explicit form is given in \eqref{optimal-F}, does not rely upon $z$.
	It follows that $\arg\min_{F\in\mathcal{F}} J(F,z)=\arg\min_{F\in\mathcal{F}} \frac{1}{r}\sum_{i=1}^{r} J(F,z_{i})$ for any initial states $z$ and $z_{i}$, $i\in\{1,\hdots,r\}$, satisfying $z_iz_i^{\prime}\succ 0$.\vspace{-4mm}
	
	Based on the definitions of $J(\cdot,\cdot)$ and $\hat{J}(\cdot,\cdot)$, an algebraic manipulation leads to\vspace{-2mm}
	\begin{align*}
	\frac{1}{r}\sum_{i=1}^{r} \hat{J}(F,v_{(i)})=& \frac{\gamma}{r} \sum_{i=1}^{r} J(F,z_{i})+\frac{1}{r} \sum_{i=1}^{r} v_{(i)}^{\prime}\Lambda v_{(i)}\\
	&+\gamma\tr[(Q+F^{\prime}RF)\sigma_w^2],
	\end{align*}
	
	\vspace{-6mm}\noindent where $z_{i}=[A \ B]v_{(i)}$.
	Since the last two terms on the right-hand side of the above equation are constants, it holds that $\arg\min_{F\in\mathcal{F}} \frac{1}{r}\sum_{i=1}^{r} J(F,z_{i})=\arg\min_{F\in\mathcal{F}} \frac{1}{ r}\sum_{i=1}^{r} \hat{J}(F,v_{(i)})$.\vspace{-4mm}
	
	From the above, it follows that $F^*=\arg\min_{F\in\mathcal{F}} J(F,z)=\arg\min_{F\in\mathcal{F}} \frac{1}{ r}\sum_{i=1}^{r} J(F,z_{i})=\arg\min_{F\in\mathcal{F}} \frac{1}{ r}\sum_{i=1}^{r} \hat{J}(F,v_{(i)})=\hat{F}^*$.
	Moreover, the uniqueness of $F^*$ guarantees that $\hat{F}^*$ is unique.
\end{proof}

\vspace{-3mm} The followings are two conclusions that will be frequently used in the following contents.\vspace{-3mm}
\begin{lemma}[Lyapunov Stability Theorems {\color{black}\cite[Chapter~3]{gu2012discrete}}] \label{Lyapunov}
	Let $A\in\mathbb{R}^{n\times n}$. Then, \vspace{-3mm}
	\begin{enumerate}
		\item if $\rho(A)<1$, then for each given matrix $M\in\mathbb{S}_{+}^n$, there exists a unique matrix $P\in\mathbb{S}_{+}^n$ satisfying $A^{\prime}PA+M=P$;
		\item $\rho(A)<1$ if and only if for each given matrix $M\in\mathbb{S}_{++}^n$, there exists a unique matrix $P\in\mathbb{S}_{++}^n$ such that $A^{\prime}PA+M=P$. 
	\end{enumerate}
\end{lemma}\vspace{-2mm}

\subsection{Primal-Dual Formulation}
\vspace{-2mm}
From the previous subsection, we know that Problem~\ref{LQG-problem2} is a non-convex optimization problem with the objective function $\frac{1}{r}\sum_{i=1}^r \hat{J}(F,v_{(i)})$, the static constraint $F\in\mathcal{F}$, and the dynamic constraint \eqref{LTI-syst}.
In this subsection, we propose an equivalent non-convex optimization formulation of Problem~\ref{LQG-problem2}, and show that its dual gap is zero.\vspace{-4mm}

We first give the definition of the non-convex optimization formulation.\vspace{-3mm}
\begin{primal-pro}\label{primal-1}
	Solve \vspace{-2mm}
	\begin{align}
	J_{p}\triangleq &\inf_{S\in\mathbb{S}^{n+m},F\in\mathbb{R}^{m\times n}} \tr (\Lambda S)\notag\\
	\st \;\;\;\; & S\succ 0,\\
	& \gamma A_FSA_F^{\prime}+\Gamma+\frac{\gamma}{1-\gamma}\bar{F}\sigma_w^2\bar{F}^{\prime}=S.\label{primal-1-Lyapunov}
	\end{align}
\end{primal-pro}
\vspace{-7mm} It is obvious to see that Primal Problem~\ref{primal-1} is a single-objective multi-variable optimization problem consisting of a linear objective function, a linear inequality constraint, and a quadratic equality constraint \cite{boyd2004convex}.
Hence, it is a non-convex optimization problem.\vspace{-4mm}

The following proposition shows the equivalence between Problem~\ref{LQG-problem2} and Primal Problem~\ref{primal-1}.\vspace{-3mm}
\begin{proposition}\label{equiv-primal-1}
	The optimal solution of Primal Problem~\ref{primal-1}, which is denoted by $(S_p,F_p)$, is unique.
	Furthermore, Primal Problem~\ref{primal-1} is equivalent to Problem~\ref{LQG-problem2} in the sense that $J_p=\hat{J}^*$ and $F_p=\hat{F}^*$.
\end{proposition}\vspace{-2mm}
\begin{proof}
	By the properties of trace, we rewrite the objective function of Problem~\ref{LQG-problem2} as\vspace{-2mm}
	\begin{align*}
	\frac{1}{r}\sum_{i=1}^{r}\hat{J}(F,v_{(i)})=\tr (\Lambda S)
	\end{align*}
	
	\vspace{-7mm}\noindent with \vspace{-3mm}
	\begin{align}\label{S-1}
	S\triangleq S(F)=\frac{1}{r}\sum_{i=1}^{r} \sum_{k=0}^{\infty} \gamma^k \mathbb{E}[v_k(F,v_{(i)})v_k(F,v_{(i)})^{\prime}].
	\end{align}
	
	\vspace{-7mm}\noindent From \eqref{augmented-system}, we have \vspace{-3mm}
	\begin{align}\label{v}
	v_k(F,v_{(i)})=A_F^k v_{(i)}+\sum_{j=0}^{k-1} A_F^j \bar{F} w_{k-j-1},\; k\ge 1.
	\end{align}
	
	\vspace{-7mm}\noindent Using the equation \eqref{v} and the fact that $w_k\sim \mathcal{N}(0_n,\sigma_w^2)$ is the i.i.d. random noise for each $k\ge 0$, we can obtain that for any $F\in\mathcal{F}$ and $k\ge 1$, it holds \vspace{-2mm}
	\begin{align*}
	&\mathbb{E}[v_k(F,v_{(i)}) v_k(F,v_{(i)})^{\prime}]\\
	=&\mathbb{E}\left[A_F^kv_{(i)}v_{(i)}^{\prime}(A_F^k)^{\prime}+\sum_{j=0}^{k-1} A_F^j \bar{F} w_{k-j-1}w_{k-j-1}^{\prime}\bar{F}^{\prime}(A_F^j)^{\prime} \right].
	\end{align*}
	
	\vspace{-6mm}\noindent Thus, $S$ in \eqref{S-1} becomes \vspace{-2mm}
	\begin{align}\label{S-2}
	S=&\frac{1}{r}\sum_{i=1}^r\left[\sum_{k=1}^{\infty}\gamma^k\Big[A_F^kv_{(i)}v_{(i)}^{\prime}(A_F^k)^{\prime}+\sum_{j=0}^{k-1}A_F^j\bar{F}\sigma_w^2\bar{F}^{\prime}(A_F^j)^{\prime}\Big]\right]\notag\\
	=&\sum_{k=1}^{\infty}\gamma^k\left[A_F^k\Gamma(A_F^k)^{\prime}+\sum_{j=0}^{k-1}A_F^j\bar{F}\sigma_w^2\bar{F}^{\prime}(A_F^j)^{\prime}\right].
	\end{align} 
	
	\vspace{-6mm}\noindent By an algebraic manipulation, one can obtain that \vspace{-3mm} 
	\begin{align*}
	S-\gamma A_F S A_F^{\prime}=\Gamma+\sum_{k=1}^{\infty} \gamma^k \bar{F} \sigma_w^2 \bar{F}^{\prime}=\Gamma+\frac{\gamma}{1-\gamma}\bar{F}\sigma_w^2\bar{F}^{\prime},
	\end{align*}
	
	\vspace{-7mm}\noindent which implies that $S$ satisfies the Lyapunov equation \eqref{primal-1-Lyapunov}. \vspace{-4mm}
	
	Since $\Gamma+\frac{\gamma}{1-\gamma}\bar{F}\sigma_w^2\bar{F}^{\prime}\succ 0$, it follows from Lemma~\ref{spectral} and the second statement of Lemma~\ref{Lyapunov} that $F\in\mathcal{F}$ if and only if there exists a unique $S\succ 0$ such that the constraint \eqref{primal-1-Lyapunov} holds.
	Therefore, we can replace the constraint $F\in\mathcal{F}$ in Problem~\ref{LQG-problem2} by $S\succ 0$ without changing its optimal solution.\vspace{-4mm}
	
	From the above discussions, we know that if $(S_p,F_p)$ is the optimal solution of Primal Problem~\ref{primal-1} and the corresponding optimal cost is $J_p$, then it holds that $S_p \succ 0$ and $\rho(A_{F_p})<1$.
	From Lemma~\ref{spectral}, $F_p\in\mathcal{F}$ and $F_p$ is a feasible point of Problem~\ref{LQG-problem2}.
	Thereby, $J_p$ is lower bounded by the optimal value of Problem~\ref{LQG-problem2}, i.e., $J_p\ge \hat{J}^*$.
	In addition, if $S_p$ is the unique solution of \eqref{primal-1-Lyapunov} with $F_p=\hat{F}^*\in\mathcal{F}$, then the resulting objective function of Primal Probelm~\ref{primal-1} is $J_p=\hat{J}^*$.
	Thus, we can conclude that $(S_p,F_p)$ is the solution of Primal Problem~\ref{primal-1} and $J_p=\hat{J}^*$.
	The uniqueness of $(S_p,F_p)$ can be shown by the uniqueness of $\hat{F}^*$.
	\vspace{-3mm}
\end{proof}

For any fixed $P\in\mathbb{S}^{n+m}$, $P_0\in\mathbb{S}_{+}^{n+m}$, define the Lagrangian function as \vspace{-2mm}
\begin{equation}
\begin{aligned}\label{Lagrangian-2}
L(P,P_0,&F,S)\triangleq \tr(\Lambda S)+\tr(-SP_0)\\
&+\tr\left[\Big(\gamma A_FSA_F^{\prime}+\Gamma+\frac{\gamma}{1-\gamma}\bar{F}\sigma_w^2\bar{F}^{\prime}-S\Big)P\right],
\end{aligned}
\end{equation}

\vspace{-6mm}\noindent where $P$ and $P_0$ are Lagrangian multipliers.
The Lagrangian dual function is defined as \vspace{-2mm}
\begin{align*}
d(P,P_0)\triangleq \inf_{S\in\mathbb{S}^{n+m},F\in\mathbb{R}^{m\times n}}L
(P,P_0,F,S).
\end{align*}

\vspace{-6mm}\noindent Then, the dual problem corresponding to Primal Problem~\ref{primal-1} is defined as \vspace{-2mm}
\begin{align*}
J_d \triangleq& \sup_{P\in\mathbb{S}^{n+m},P_0\in\mathbb{S}_{+}^{n+m}} d(P,P_0) \\
=&\sup_{P\in\mathbb{S}^{n+m},P_0\in\mathbb{S}_{+}^{n+m}} \inf_{S\in\mathbb{S}^{n+m},F\in\mathbb{R}^{m\times n}}L(P,P_0,F,S),
\end{align*}

\vspace{-6mm}\noindent where $L(P,P_0,F,S)$ is as defined in \eqref{Lagrangian-2}. \vspace{-4mm}

The weak duality \cite[Chapter~5]{boyd2004convex} guarantees that $J_d\le J_p$ holds, where $J_p-J_d\ge 0$ is called the duality gap.
When the duality gap is zero, we say that the optimization problem has strong duality \cite[Chapter~5]{boyd2004convex}.
The following theorem shows that Primal Problem~\ref{primal-1} has strong duality. \vspace{-3mm}

\begin{theorem}[Strong Duality]\label{thm-1}
	The strong duality holds for Primal Problem~\ref{primal-1}, that is, $J_p=J_d$.
\end{theorem}\vspace{-3mm}
\begin{proof}
	See Appendix~\ref{proof_thm1}.
\end{proof}\vspace{-2mm}


\subsection{MB-PD Algorithm}
\vspace{-2mm}
It is well-known that for a non-convex optimization problem, if its objective and constraints functions are differentiable and it has strong duality, then any primal and dual optimal points must satisfy the corresponding KKT conditions, but not vise versa\cite[Chapter~5.5.3]{boyd2004convex}.
However, if the solution to the KKT conditions is unique, then this solution must be the primal and dual optimal points.
In view of this, we first derive the KKT conditions and show the uniqueness of their solutions in this subsection, and then design an algorithm to seek the solution to Problem~\ref{LQG-problem2} according to the solutions to the KKT conditions.\vspace{-3mm}

\begin{lemma}\label{KKT}
	Suppose that $(S_p,F_p)$ is the primal optimal point and $(P_p,P_{0,p})$ is the dual optimal point of Primal Problem~\ref{primal-1}.
	Then, $(S_p,F_p,P_p,P_{0,p})$ satisfies the KKT conditions for $(S,F,P,P_0)$\vspace{-2mm}
	\begin{align}
	&\gamma A_{F}SA_F^{\prime}+\Gamma+\frac{\gamma}{1-\gamma}\bar{F}\sigma_w^2\bar{F}^{\prime}-S=0,\label{primal-feasibility1}\\
	&S\succ 0,\label{primal-feasibility2}\\
	&P_0=\textbf{0}_{n+m},\label{dual-feasibility}\\
	&\gamma A_F^{\prime}PA_F+\Lambda=P,\label{stationary1}\\
	&2(P_{12}^{\prime}+P_{22}F)\Big(\gamma\left[A\;\;B\right]S\left[A\;\;B\right]^{\prime}+\frac{\gamma}{1-\gamma}\sigma_w^2\Big)=0.\label{stationary2}
	\end{align}
\end{lemma}\vspace{-7mm}
 \begin{proof}
	From the conclusions in \cite[Chapter~5.9.2]{boyd2004convex}, the KKT conditions of Primal Problem~\ref{primal-1} are as follows:\vspace{-4mm}
	\begin{enumerate}
		\item primal feasibility condition: \vspace{-2mm}
		\begin{align*}
		&\gamma A_FSA_F^{\prime}+\Gamma+\frac{\gamma}{1-\gamma}\bar{F}\sigma_w^2\bar{F}^{\prime}-S=0,\\
		&S \succ 0;
		\end{align*}
		
		\item \vspace{-3mm} \noindent dual feasibility condition:\vspace{-2mm}
		\begin{align*}
		P_0\succeq 0;
		\end{align*}
		\item \vspace{-3mm} \noindent complementary slackness condition:\vspace{-2mm}
		\begin{align*}
		\tr(SP_0)=0;
		\end{align*}
		\item \vspace{-3mm} \noindent stationary conditions:\vspace{-2mm}
		\begin{align*}
		&\!\frac{\partial L}{\partial S}=\Lambda+\gamma A_F^{\prime}SA_F-P-P_0=0,\\
		&\!\frac{\partial L}{\partial F}=2(P_{12}^{\prime}+P_{22}F)\Big(\gamma\left[A\;\;B\right]S\left[A\;\;B\right]^{\prime}\!+\!\frac{\gamma}{1-\gamma}\sigma_w^2\Big)=0.	
		\end{align*}
	\end{enumerate}\vspace{-2mm}
	By the second statement of Lemma~\ref{Lyapunov}, $\Gamma+\frac{\gamma}{1-\gamma}\sigma_w^2\succ 0$ and $F\in\mathcal{F}$ guarantee that $S\succ 0$, which further implies that the only solution to the equation $\tr(SP_0)=0$ is $P_0=\textbf{0}_{m+n}$.
	Hence, the KKT conditions in \eqref{primal-feasibility1}-\eqref{stationary2} can be obtained.
	According to \cite[Chapter~5.5.3]{boyd2004convex}, the strong duality guarantees that the solutions to the KKT conditions must be the primal and dual optimal points.
\end{proof}

\vspace{-3mm}The following lemma is given to show that the uniqueness of the solutions to the KKT conditions \eqref{primal-feasibility1}-\eqref{stationary2}, which guarantees that the solutions satisfying the KKT conditions must be the primal and dual optimal points.\vspace{-3mm}
\begin{lemma}
	Suppose that $(S,F,P,P_{0})$ satisfies the KKT conditions \eqref{primal-feasibility1}-\eqref{stationary2}.
	Then, $(S,F,P,P_{0})$ is unique.
	Furthermore, it holds that $F=F_p$, $P=P_p$, $P_{0}=\textbf{0}_{m+n}$, and $S_p$ is the unique solution to \eqref{primal-feasibility1} with $F=F_p$ and $P=P_p$.
\end{lemma}\vspace{-3mm}

\begin{proof}
	By observation, we can know that the sufficient condition for \eqref{stationary2} is $F=-(P_{22})^{-1}P_{12}^{\prime}$.
	To show this lemma, we first show that $(F,P)$ satisfying $\gamma A_F^{\prime}PA_F+\Lambda=P$,  $F=-(P_{22})^{-1}P_{12}^{\prime}$, and $P\succeq 0$ is exactly $(F_p,P_p)$ with $F_p\in\mathcal{F}$.
	Using the same argument as in the proof of Lemma~\ref{primal-2-prop}, we have $P_p=P^*$.
	From Proposition~\ref{equiv-F} and Proposition~\ref{equiv-primal-1}, it holds that $F_p=F^*$.
	In view of this, we only need to show that $(F,P)$ satisfying $\gamma A_F^{\prime}PA_F+\Lambda=P$,  $F=-(P_{22})^{-1}P_{12}^{\prime}$, and $P\succeq 0$ is exactly $(F^*,P^*)$.\vspace{-4mm}
	
	By substituting $F=-(P_{22})^{-1}P_{12}^{\prime}$ into \eqref{stationary1}, it follows that\vspace{-3mm}
	\begin{align*}
		\gamma \left[A \ \ B \right]^{\prime}(P_{11}-P_{12}P_{22}^{-1}P_{12}^{\prime})\left[	A \ \ B	\right]+\Lambda=P.
	\end{align*}
	
	\vspace{-6mm}\noindent Define $\bar{X}\triangleq P_{11}-P_{12}P_{22}^{-1}P_{12}^{\prime}$.
	The above equation becomes
	$\gamma \left[A \ B \right]^{\prime}\bar{X}\left[A \ B\right]+\Lambda=P$, the expansion of which can be written as\vspace{-2mm}
	\begin{align*}
		\left[
		\begin{array}{cc}
		\gamma A^{\prime}\bar{X}A+Q & \gamma A^{\prime}\bar{X} B\\
		\gamma B^{\prime}\bar{X}A   & \gamma B^{\prime}\bar{X}B+R
		\end{array}
		\right]=\left[
		\begin{array}{cc}
		P_{11} & P_{12}\\
		P_{12}^{\prime} & P_{22}
		\end{array}
		\right],
	\end{align*}
	
	\vspace{-6mm}\noindent i.e., $P_{11}=\gamma A^{\prime}\bar{X}A+Q$, $P_{12}=\gamma A^{\prime}\bar{X} B$, and $P_{22}=\gamma B^{\prime}\bar{X}B+R$.
	Substituting these expressions into the definition $\bar{X}\triangleq P_{11}-P_{12}P_{22}^{-1}P_{12}^{\prime}$ yields that\vspace{-2mm}
	\begin{align*}
		Q+\gamma A^{\prime}\bar{X}A-\gamma^2 A^{\prime}\bar{X} B(R+\gamma B^{\prime}\bar{X}B)^{-1}B^{\prime}\bar{X}A=\bar{X}.
	\end{align*}
	
	\vspace{-6mm}\noindent This is exactly the ARE \eqref{ARE}, which has a unique solution.
	Therefore, it holds that $\bar{X}=X^*$.
	From the definitions \eqref{optimal-F} and \eqref{P}, we have $F=F^*$ and $P=P^*$.\vspace{-4mm}
	
	Then, based on the fact that $F_p=F^*\in\mathcal{F}$ and $\Lambda\succeq 0$,  the first statement of Lemma~\ref{Lyapunov} shows that there exists a unique $S=S_p$ satisfying \eqref{primal-feasibility1} with $F=F_p$ and $P=P_p$.
\end{proof}\vspace{-0mm}

\vspace{-3mm}Note that to calculate the primal optimal point $F_p$, one needs to know $P_p$, which is uniquely determined by solving \eqref{stationary1}.
However, \eqref{stationary1} is dependent upon the primal optimal point $F_p$, which means that the stationary conditions  \eqref{stationary1} and \eqref{stationary2} are coupled with each other, so that $F_p$ and $P_p$ cannot be obtained independently. 
In what follows, we propose an algorithm, cf. Algorithm~\ref{alg1}, to iteratively solve \eqref{stationary1} and \eqref{stationary2}.
Note that once $F_p$ and $P_p$ are attained, one can uniquely determine the primal variable $S_p$ by solving \eqref{primal-feasibility1}. 

\begin{algorithm}[t]
	\caption{MB-PD Algorithm}  
	\label{alg1}
	\begin{algorithmic}[1]
		\STATE \textbf{Initialization:} $F_p^0\in\mathcal{F}$, $\epsilon>0$, and set $s=0$;
		\STATE \textbf{Repeat}
		\STATE \quad  \textbf{dual update:} solve $P^s$ from \vspace{-2mm}
		$$\gamma (A_{F_p^s})^{\prime}P_p^sA_{F_p^s}+\Lambda=P_p^s;$$		\\
		\STATE \vspace{-2mm} \quad \textbf{primal update:} $F_p^{s+1}=-(P_{p,22}^s)^{-1}(P_{p,12}^s)^{\prime}$;
		\STATE \quad $s\leftarrow s+1;$
		\STATE \textbf{Until} $\|F_p^{s}-F_p^{s-1}\|\le\epsilon$;
		\STATE \textbf{Return} $F_p^s$.
	\end{algorithmic}
\end{algorithm}

\vspace{-4mm}We have the following convergence result:\vspace{-3mm}
	\begin{lemma}
		Given an initial stabilizing control gain $F_p^0$. 
		Consider the two sequences $\{P_p^s\}_{s=0}^{\infty}$ and $\{F_p^s\}_{s=0}^{\infty}$ obtained from Algorithm~\ref{alg1}.
		We have the following properties:\vspace{-3mm}
		\begin{enumerate}
			\item $\{P_p^s\}_{s=0}^{\infty}$ is a non-increasing sequence, that is, $P_p^s \succeq P_p^{s+1}$;
			\item $F_p^s$ obtained at each iteration step is stabilizing, that is, $F_p^s\in\mathcal{F}$ for each $s$;
			\item $\lim_{s\to\infty}P_p^s=P_p=P^*$ and $\lim_{s\to\infty}F_p^s=F_p=F^*$, where $P^*$ and $F^*$ are defined in \eqref{P} and \eqref{optimal-F}, respectively.
		\end{enumerate}
	\end{lemma}\vspace{-2mm}
	\begin{proof}
		Note that Algorithm~\ref{alg1} can be transformed into Algorithm~1 developed in \cite{lee2018primal} by replacing $(A,B)$ with $(\gamma^{1/2}A,\gamma^{1/2}B)$.
		Therefore, this lemma can be shown directly by using the proof of Proposition~8 in \cite{lee2018primal} by some algebraic manipulations.
	\end{proof}\vspace{-2mm}

\subsection{Model-Free Implementation of Algorithm~\ref{alg1}}
\vspace{-2mm}
From the equation \eqref{S-1} that is defined in the proof of Proposition~\ref{equiv-primal-1}, $S(F_p^s)$ can be equivalently written as \vspace{-2mm}
\begin{align}
S(F_p^s)=\frac{1}{r}\sum_{i=1}^{r}\sum_{k=0}^{\infty} \gamma^k \mathbb{E}\left[v_k\big(F_p^s,v_{(i)}\big)v_k\big(F_p^s,v_{(i)}\big)^{\prime}\right].
\end{align}

\vspace{-6mm}\noindent We truncate the time horizon by $K>0$, and define\vspace{-2mm}
\begin{align}
\tilde{S}(F_p^s)=\frac{1}{r}\sum_{i=1}^{r}\sum_{k=0}^{K} \gamma^k \mathbb{E}\left[v_k\big(F_p^s,v_{(i)}\big)v_k\big(F_p^s,v_{(i)}\big)^{\prime}\right].
\end{align}

\vspace{-6mm}\noindent Note that for any $K>0$, it holds that $\tilde{S}(F_p^s)\succ 0$ due to the positive definiteness of $\frac{1}{r}\sum_{i=1}^r v_{(i)}v_{(i)}^{\prime}$.
In view of this, the stationary condition \eqref{stationary1} holds if and only if $\tilde{S}(F_p^s)^{\prime}\Big(\gamma A_{F_p^s}^{\prime}P_p^sA_{F_p^s}+\Lambda-P_p^s\Big)\tilde{S}(F_p^s)=0$, which can be rewritten as the following linear matrix equation\vspace{-2mm}
\begin{align}\label{compute-P}
\tilde{S}(F_p^s)^{\prime}P_p^s\tilde{S}(F_p^s)-\gamma \tilde{W}(F_p^s)^{\prime}P_p^s\tilde{W}(F_p^s)=\tilde{S}(F_p^s)^{\prime}\Lambda\tilde{S}(F_p^s),
\end{align}

\vspace{-8mm}\noindent where \vspace{-2mm}
\begin{align}
\tilde{W}(F_p^s)=\frac{1}{r}\sum_{i=1}^r\sum_{k=0}^{K} \gamma^k \mathbb{E}\left[v_{k+1}\big(F_p^s,v_{(i)}\big)v_k\big(F_p^s,v_{(i)}\big)^{\prime}\right].
\end{align}

\vspace{-7mm}\noindent Therefore, for a fixed and stabilizing control gain $F_p^s$, the corresponding dual variable $P_p^s$ can be obtained by solving the linear matrix equation \eqref{compute-P}.\vspace{-4mm}

Based on the above analysis, we give a detailed model-free implementation of the proposed MB-PD algorithm, and summarize it in Algorithm~\ref{alg2}.
Similar to Algorithm~\ref{alg-imp-off-policy}, the mathematical expectations are approximated by the numerical averages herein.

\begin{algorithm}[t]
	\caption{Model-Free Implementation of MB-PD Algorithm}  
	\label{alg2}
	\begin{algorithmic}[1]
		\STATE \textbf{Initialization:} set a stabilizing control gain $F_p^{0}$, $s=0$, $K>0$, and $\epsilon>0$;
		\STATE \textbf{Repeat}
		\STATE \quad \textbf{primal update:} $\tilde{S}=0$, $\tilde{W}=0$;
		\STATE \quad \textbf{for} $i=1:r$
		\STATE \qquad $\bar{S}=0$, $\bar{W}=0$;
		\STATE \qquad \textbf{for} $q=1:N$
		\STATE \qquad \quad apply $u_k=F_p^sx_k$ for $K$ time steps;
		\STATE \qquad \quad sample $\{x_k,u_k\}_{k=0,\hdots,K}$;
		\STATE \qquad \quad calculate $S=\sum_{k=0}^K \gamma^k v_k v_k^{\prime}$ with $v_k=[x_k^{\prime} \ u_k^{\prime}]^{\prime}$;
		\STATE \qquad \quad calculate $W=\sum_{k=0}^K \gamma^k v_{k+1} v_{k}^{\prime}$;
		\STATE \qquad \quad $\bar{S}=\bar{S}+S$, $\bar{W}=\bar{W}+W$ ;
		\STATE \qquad \textbf{End}
		\STATE \quad $\bar{S}=\bar{S}/N$, $\bar{W}=\bar{W}/N$;
		\STATE \quad $\tilde{S}=\tilde{S}+\bar{S}$, $\tilde{W}=\tilde{W}+\bar{W}$;
		\STATE \quad \textbf{End}
		\STATE \quad \textbf{dual update:}
		 calculate $P_p^s$ by solving \vspace{-3mm}
		$$\tilde{S}^{\prime}P_p^s\tilde{S}-\gamma \tilde{W}^{\prime}P_p^s\tilde{W}=\tilde{S}^{\prime}\Lambda\tilde{S};$$ 
		\STATE \vspace{-3mm} \quad \textbf{primal update:} update control gain from \vspace{-3mm}
		$$F_p^{s+1}=-(P_{p,22}^s)^{-1}(P_{p,12}^s)^{\prime};$$		
		\STATE \vspace{-3mm} \quad $s\leftarrow s+1$;
		\STATE \textbf{Until} $\|F_p^{s}-F_p^{s-1}\|\le\epsilon$;
		\STATE \textbf{Return} $F_p^s$.
	\end{algorithmic}
\end{algorithm}

\vspace{-4mm}Note that different from the Algorithm~\ref{alg-imp-off-policy}, the primal-dual optimization-based algorithm developed in this section has the  advantage of averting the requirement of the excitation signal, which may cause undesirable system oscillation.
However, this is at the cost of the requirement of more data.
Specifically, it is assumed that we can sample the state-input trajectories generated under a certain set of initial vectors which are linearly independent in $\mathbb{R}^n$.\vspace{-2mm}

\section{Equivalency Between PI and MB-PD Algorithms}\label{alg-equiv}\vspace{-2mm}

In the previous two sections, we solve the stochastic LQR problem from the perspective of RL and primal-dual optimization, respectively.
The goal of this section is to show the equivalency of the algorithms developed before.
Specifically, we show that classical PI and MB-PD algorithms are equivalent in the sense that there is a corresponding relation in the involved iterations.
The specific relationship is described in the following theorem.\vspace{-3mm}

\begin{theorem}\label{thm3}
	The dual and primal update steps in Algorithm~\ref{alg1} can be interpreted as the policy evaluation and policy improvement steps in  Algorithm~\ref{alg-offline}, respectively, in the sense that $X^s=(\bar{F}_p^s)^{\prime}P_p^s \bar{F}_p^s$ and $F_p^{s}=F^{s}$ with $\bar{F}_p^s=[I_n \ (F_p^s)^{\prime}]^{\prime}$.
\end{theorem}\vspace{-2mm}
\begin{proof}
	By pre- and post-multiplying $(A_{F_p^s})^{\prime}P_p^sA_{F_p^s}+\Lambda=P_p^s$ by $[I_n \ \ (F_p^s)^{\prime}]$ and its transpose, one can obtain that \vspace{-2mm}
	\begin{equation}
	\begin{aligned}
	&\gamma (A+BF_p^s)^{\prime} (\bar{F}_p^s)^{\prime}P_p^s\bar{F}_p^s (A+B F_p^s)+Q+(F_p^s)^{\prime}R F_p^s\\ 
	=&(\bar{F}_p^s)^{\prime}P_p^s\bar{F}_p^s.\label{equiv-1}
	\end{aligned}
	\end{equation}
	
	\vspace{-6mm}\noindent Define $X_p^s\triangleq(\bar{F}_p^s)^{\prime}P_p^s\bar{F}_p^s$.
	Then, \eqref{equiv-1} becomes \vspace{-2mm}
	\begin{align*}
	\gamma (A+B F_p^s)^{\prime}X_p^s(A+BF_p^s)+Q+(F_p^s)^{\prime}R F_p^s
	=X_p^s,
	\end{align*}
	
	\vspace{-6mm}\noindent which has a unique solution since $F_p^s\in\mathcal{F}$ and $Q+(F_p^s)^{\prime}R F_p^s$ is positive semidefinite.
	By observation, we know that the dual update step in Algorithm~\ref{alg1} is equivalent to the policy evaluation step in Algorithm~\ref{alg-offline} in the sense that $X^s=X_p^s=(\bar{F}_p^s)^{\prime}P_p^s\bar{F}_p^s$. \vspace{-4mm}
	
	Let $P_p^s=\left[
	\begin{array}{cc}
	P_{p,11}^s & P_{p,12}^s\\
	P_{p,21}^s & P_{p,22}^s
	\end{array}
	\right]$
	with $P_{p,11}^s\in\mathbb{R}^{n\times n}$, $P_{p,22}^s\in\mathbb{R}^{m\times m}$, and $P_{p,12}^s=(P_{p,21}^s)^{\prime}\in\mathbb{R}^{n\times m}$.
	Expanding $(\bar{F}_p^s)^{\prime}P_p^s\bar{F}_p^s$ yields that\vspace{-2mm}
	\begin{equation}
	\begin{aligned}\label{equiv-2}
	P_{p,11}^s+(F_p^s)^{\prime}P_{p,21}^s+P_{p,12}^sF_p^s+(F_p^s)^{\prime}P_{p,22}^sF_p^s=X^s.
	\end{aligned}
	\end{equation}
	
	\vspace{-6mm}\noindent Expanding the left-hand side of \eqref{-PE} yields that \vspace{-2mm}
	\begin{equation}
	\begin{aligned}\label{equiv-3}
	(Q+\gamma A^{\prime}X^sA)+(F^s)^{\prime}&(\gamma B^{\prime}X^s A)+\gamma A^{\prime}X^sBF^s\\
	&+(F^s)^{\prime}(R+\gamma B^{\prime}X^sB)F^s=X^s.
	\end{aligned}
	\end{equation}
	
	\vspace{-7mm}\noindent Making the corresponding terms in \eqref{equiv-2} and \eqref{equiv-3} identical, it follows that $P_{p,11}^s=Q+\gamma A^{\prime}X^s A$, $P_{p,12}^s=\gamma A^{\prime}X^sB$, and $P_{p,22}^s=R+\gamma B^{\prime}X^sB$, 
	which further imply that
	$F_p^{s+1}=F^{s+1}$.
\end{proof}\vspace{-2mm}

\begin{example}\rm
	Considering the following discrete-time linear system:\vspace{-2mm}
	\begin{align*}
	x_{k+1}=2x_k+u_k+w_k.
	\end{align*}
	
	\vspace{-6mm}\noindent The weight matrices and discount factor are selected as $Q=1$, $R=1$, and $\gamma=0.7$.
	The system noise is assumed to be generated according to the Gaussian distribution $\mathcal{N}(0,1)$.
	We choose $F^0=F_p^0=-1$ as the initial control gain matrix.
	Suppose that the system matrices $A$ and $B$ are known.\vspace{-4mm}
	
	By implementing the classical PI algorithm, i.e., Algorithm~\ref{alg-offline}, for $4$ steps, one can obtain that \vspace{-2mm}
	\begin{align*}
	X^1&=6.6666,\ \  X^2=4.0675, \ \ X^3=3.9353,\ \ X^4=3.9345;\\
	F^1&=-1.6471, F^2=-1.4801, F^3=-1.4673, F^4=-1.4673.
	\end{align*}
	
	\vspace{-6mm}Then we implement the MB-PD algorithm, i.e., Algorithm~\ref{alg1}, for $4$ steps, producing that \vspace{-2mm}
	\begin{align*}
	P_p^1&=\left[
	\begin{array}{cc}
	19.6666 & 9.3333\\
	9.3333  & 5.6667
	\end{array}
	\right], \ \
	P_p^2=\left[
	\begin{array}{cc}
	12.3889 & 5.6945\\
	5.6945  & 3.8472
	\end{array}
	\right],\\
	P_p^3&=\left[
	\begin{array}{cc}
	12.0188 & 5.5094\\
	5.5094  & 3.7547
	\end{array}
	\right], \ \
	P_p^4=\left[
	\begin{array}{cc}
	12.0116 & 5.5083\\
	5.5083  & 3.7542
	\end{array}
	\right];
	\end{align*}
	\vspace{-7mm}
	\begin{align*}
	F_p^1=-1.6471, F_p^2=-1.4801, F_p^3=-1.4673, F_p^4=-1.4673.
	\end{align*}
	
	\vspace{-6mm}\noindent From $X_p^s=(\bar{F}_p^s)^{\prime}P_p^s\bar{F}_p^s$, it follows that \vspace{-2mm}
	\begin{align*}
	X_p^1=6.6667,\ \ X_p^2=4.0675,\ \ X_p^3=3.9353,\ \ X_p^4=3.9345.
	\end{align*}
	
	\vspace{-6mm}By simple observation and calculation, we can verify the correctness of the conclusion given in Theorem~\ref{thm3}.

\end{example}
\vspace{-2mm}

\section{Simulations}\label{simu}
\vspace{-2mm}
In this section, we provide a numerical example to evaluate the performance of the proposed model-free algorithms.\vspace{-4mm}

Considering the following discrete-time linear system:\vspace{-2mm}
\begin{align*}
x_{k+1}=\left[
\begin{array}{cc}
0.5  &  1\\
0.25 &  0.5
\end{array}\right]x_k+\left[
\begin{array}{c}
1\\
1
\end{array}\right]u_k+w_k.
\end{align*}

\vspace{-6mm}\noindent The weight matrices and discount factor are selected as $Q=\textbf{I}_2$, $R=1$, and $\gamma=0.7$.
The system noise $w_k$ is assumed to be generated according to the Gaussian distribution $\mathcal{N}(0_2,\textbf{I}_2)$.
When one knows the exact system matrices $A$ and $B$, the optimal control gain $F^*$ can be obtained by solving the ARE \eqref{ARE}.
The result is $F^s=\left[ -0.2446 \ \  -0.4892\right]$.\vspace{-2mm}

\subsection{Performance of Algorithm~\ref{alg-imp-off-policy}}
\vspace{-2mm}
When implement Algorithm~\ref{alg-imp-off-policy}, the initial state $z$ is randomly generated. 
Inspired by \cite{kiumarsi2017h}, the probing noise is considered as \vspace{-2mm}
\begin{align*}
e_k=0.2\sin(1.009k)+\cos^2(0.538k)+\sin(0.9k)+\cos(100k).
\end{align*}

\begin{figure}[h]
	\centering
	\vspace{-0cm} 
	\includegraphics[width=8.3cm]{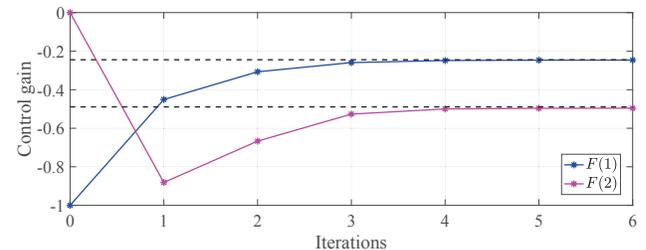}
	\vspace{-4mm}
	\caption{Values of $F^{s}$ at each iteration step. The black dotted lines are baselines for two components of the optimal control gain $F^*$.}
	\label{fig:alg3-conv}
\end{figure}

\vspace{-6mm} Initial stabilizing control gain is chosen as $F^0=[-1 \ 0]$. We use the numerical average of $N=15$ trajectories to approximate the mathematical expectation, and sample $K=20$ data points at each trajectory for the BLS method.
The tolerant error is $\varepsilon=10^{-3}$.
We can observe from Fig.~\ref{fig:alg3-conv} that Algorithm~\ref{alg-imp-off-policy} stops after $6$ iterations and returns the estimated control gain $F^6=\left[-0.2464  \ \  -0.4960\right]$.

\begin{figure}
	\centering
	\vspace{-0cm} 
	\subfigtopskip=0pt 
	\subfigbottomskip=0pt 
	\subfigcapskip=-2pt 
	\subfigure[\hspace{-0.5cm}]
	{
		\includegraphics[width=3.8cm]{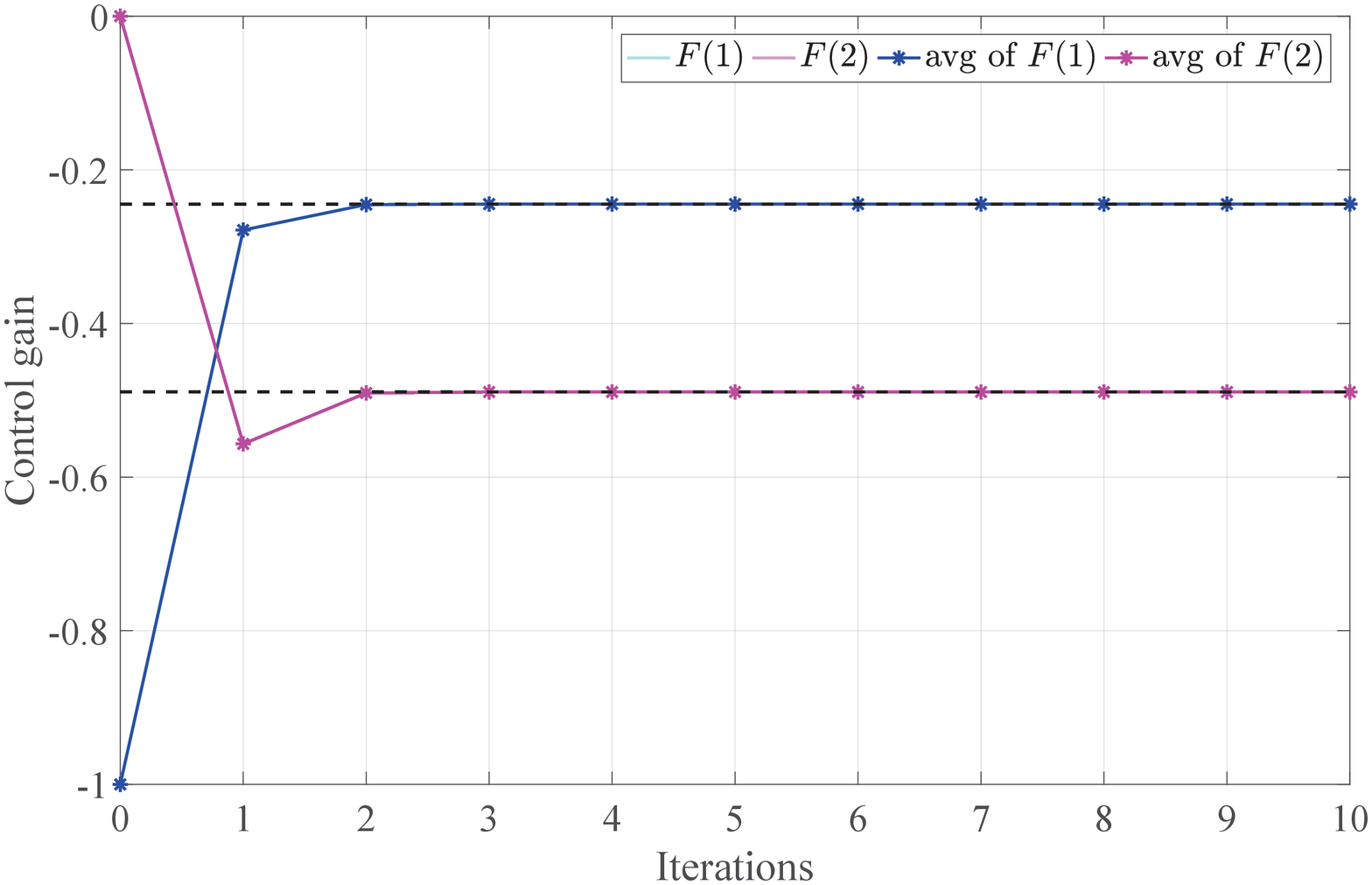}
		\label{fig:alg3-sigma0}
	}
	\subfigure[\hspace{-0.5cm}]
	{
		\includegraphics[width=3.8cm]{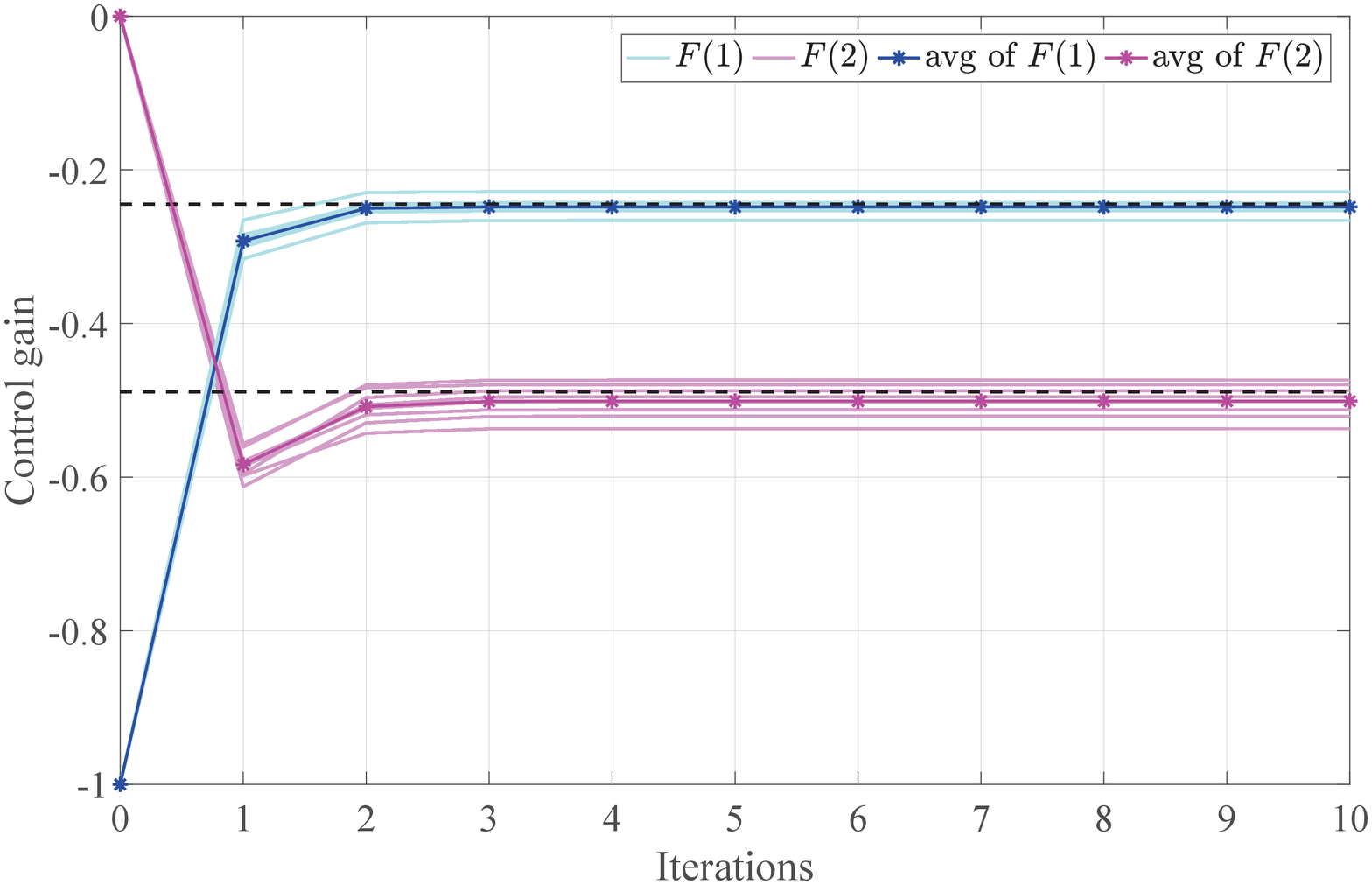}
		\label{fig:alg3-sigma01}
	}
	\subfigure[\hspace{-0.5cm}]
	{
		\includegraphics[width=3.8cm]{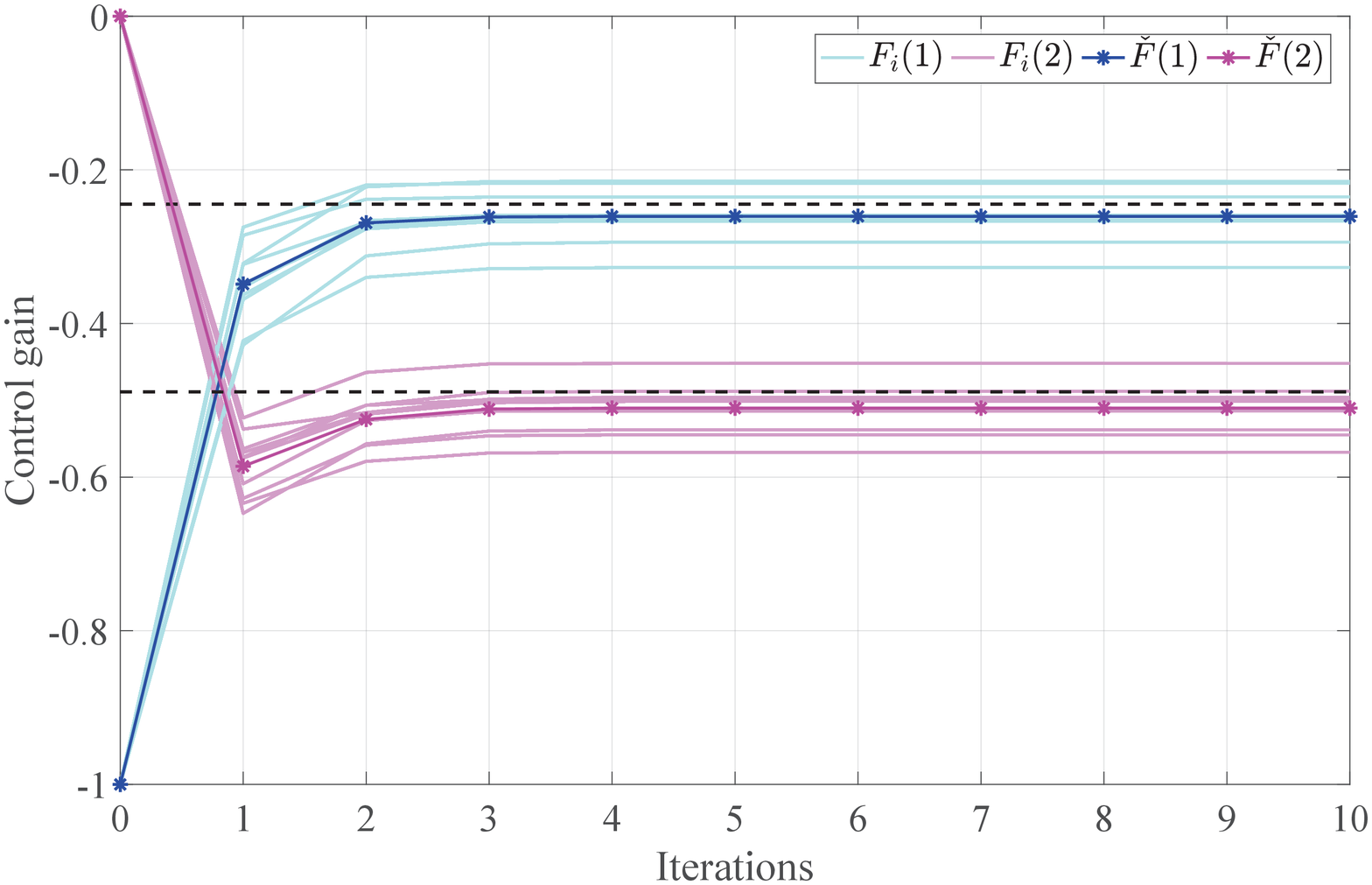}
		\label{fig:alg3-sigma05}
	}
	\subfigure[\hspace{-0.5cm}]
	{
		\includegraphics[width=3.8cm]{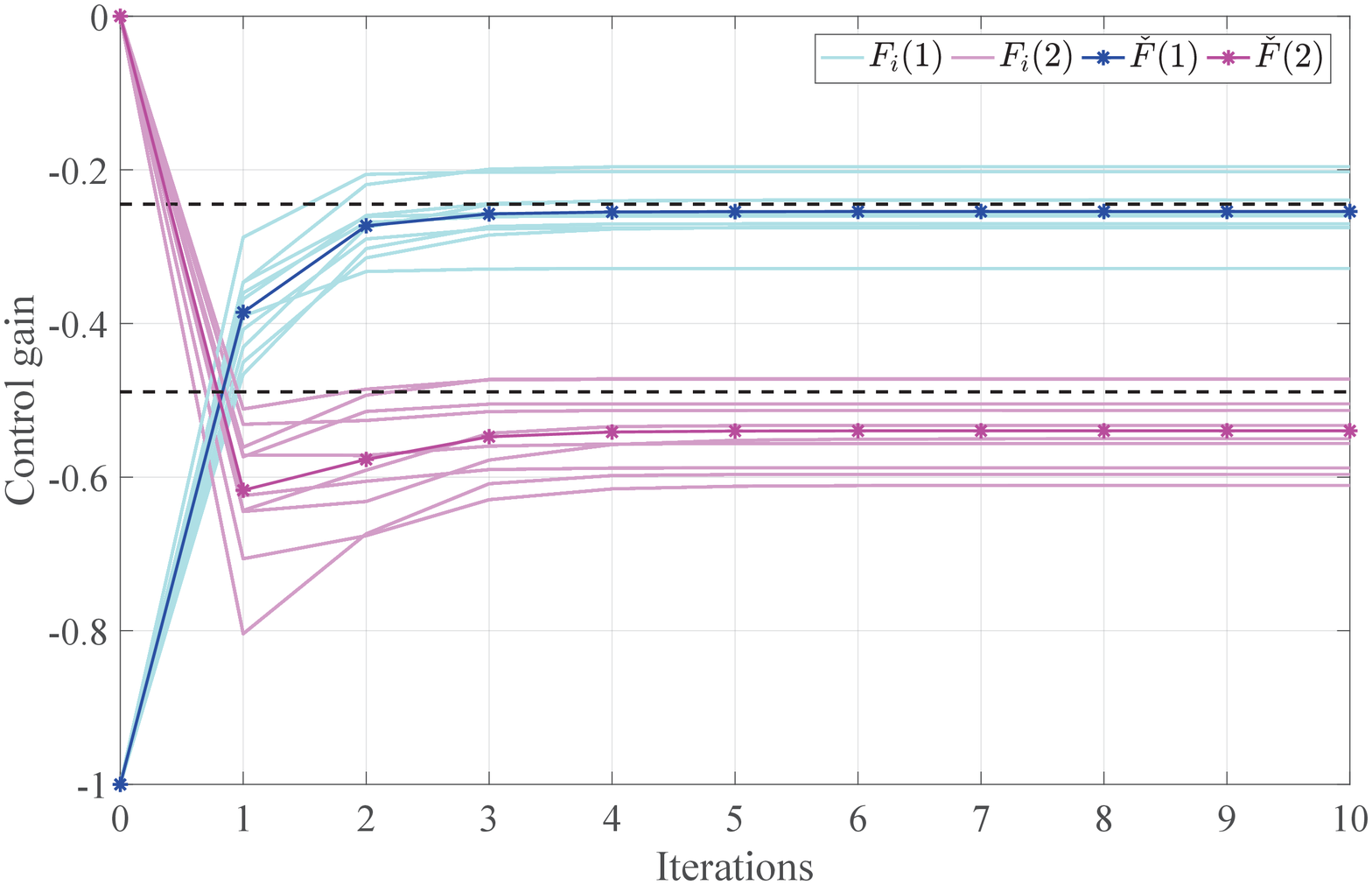}
		\label{fig:alg3-sigma1}
	}
	\caption{Values of $F^{s}$ at each experiment and the resulting average values $\check{F}^s$. The black dotted lines are baselines for two components of the optimal control gain $F^*$. (a) $\sigma_w^2=\textbf{0}_2$. (b) $\sigma_w^2=0.1\textbf{I}_2$. (c) $\sigma_w^2=0.5\textbf{I}_2$. (d) $\sigma_w^2=\textbf{I}_2$.}
	\label{fig:alg3-sigma}
\end{figure}

\begin{figure}[h]
	\centering
	\vspace{-0cm} 
	\setlength{\abovecaptionskip}{-0.1cm}
	\setlength{\belowcaptionskip}{0pt}
	\includegraphics[width=8.3cm]{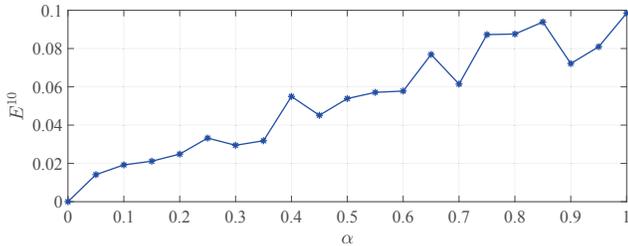}
	\vspace{-4mm}
	\caption{Average approximation errors of $Y=10$ experiments at different values of $\alpha$, where the variance of the system noise is $\sigma_w^2=\alpha \textbf{I}_2$.}
	\label{fig:alg3-sigma-trend}
\end{figure}

Considering that the system noise is randomly generated, we use Monte-Carlo method to validate the effectiveness of Algorithm~\ref{alg-imp-off-policy}, as well as to show the influence of the variance of the system noise.
We implement Algorithm~\ref{alg-imp-off-policy} for $Y=10$ times, and each experiment is stopped when $s=10$.
The average value of the resulting control gain from $Y$ experiments is denoted by $\check{F}^s=\frac{1}{Y}\sum_{i=1}^{Y}F_i^s$, where $F_i^s$ is the control gain obtained at the $s$-th step from the $i$-th experiment.
We use $\text{E}^s=\frac{1}{Y}\sum_{i=1}^Y \|F_i^s-F^*\|$ to denote the average approximation error of the $s$-th iteration step from $Y$ experiments.
Fig.~\ref{fig:alg3-sigma} shows the value of the control gain $F^s$ at each experiment as well as the resulting average gain $\check{F}^s$, when the variance of the system noise is set as $\sigma_w^2=\alpha \textbf{I}_2$ with $\alpha$ being $0$, $0.1$, $0.5$, and $1$, respectively.
Note that $\text{E}^{10}=\frac{1}{L}\sum_{i=1}^L \|F_i^{10}-F^*\|$ denotes the average approximation error on convergence, that is, the average deviation of the obtained control gain at the $i$-th experiment from the optimal control gain $F^*$.
Fig.~\ref{fig:alg3-sigma-trend} shows the variation of the average approximation error $\text{E}^{10}$, when $\alpha$ is increased from $0$ to $1$.
From Fig.~\ref{fig:alg3-sigma} and Fig.~\ref{fig:alg3-sigma-trend}, we conclude that roughly speaking, the resulting approximation errors are increased as the increase of the variance of the system noise.
That is, a stronger system noise may lead to a higher control gain approximation error.
Moreover, when $\sigma_w^2=\textbf{0}_2$, $F^s$ obtained from Algorithm~\ref{alg-imp-off-policy} converges to the optimal feedback gain $F^*$.


\subsection{Performance of Algorithm~\ref{alg2}}
\begin{figure}[h]
	\centering
	\vspace{-0cm} 
	\setlength{\abovecaptionskip}{-0.1cm}
	\setlength{\belowcaptionskip}{0pt}
	\includegraphics[width=8.3cm]{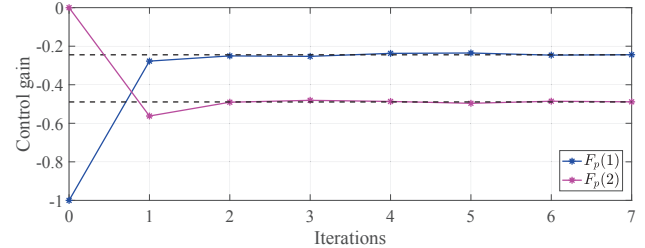}
	\vspace{-4mm}
	\caption{Values of $F_p^{s}$ at each iteration step. The black dotted lines are baselines for two components of the optimal control gain $F^*$.}
	\label{fig:alg5-conv}
\end{figure}

\vspace{-2mm}
When implement Algorithm~\ref{alg2}, we truncate the time horizon by $K=10$ and let $r=3$.
Three pairs of initial states and initial control inputs are  $z_{(1)}=[-1 \ \ 3]$, $u_{(1)}=-2$, $z_{(2)}=[2 \ \ -1]$, $u_{(2)}=-5$, $z_{(3)}=[-3 \ \ 3]$, $u_{(3)}=-8$. 
The initial stabilizing control gain is chosen as $F_p^0=[-1 \ 0]$.
We use the numerical average of $N=15$ trajectories to approximate the mathematical expectation.
The desired convergence precision is set as $\varepsilon=5\times10^{-3}$.
We can observe from Fig.~\ref{fig:alg5-conv} that Algorithm~\ref{alg2} stops after $7$ iterations and returns the estimated control gain $F_p^7=\left[ -0.2443 \ \  -0.4884\right]$.\vspace{-2mm}

\begin{figure}[h]
	\centering
	\vspace{-0cm} 
	\setlength{\abovecaptionskip}{-0.1cm}
	\setlength{\belowcaptionskip}{0pt}
	\includegraphics[width=8.3cm]{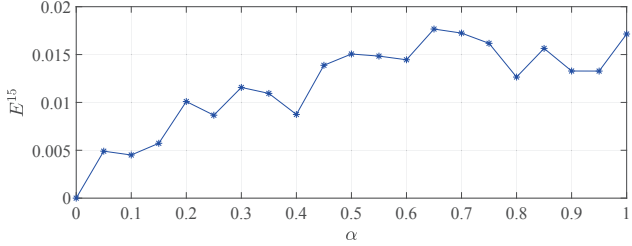}
	\vspace{-4mm}
	\caption{Average approximation errors of $Y=10$ experiments at different values of $\alpha$, where the variance of the system noise is $\sigma_w^2=\alpha \textbf{I}_2$.}
	\label{fig:alg5-sigma-ave-trend}
\end{figure}

Similar to the previous subsection, we use Monte-Carlo method to discuss the influence of the system noise herein.
We implement Algorithm~\ref{alg2} for $Y=10$ times, and stop each experiment when $s=15$.
We use $\check{F}_p^s=\frac{1}{Y}\sum_{i=1}^{Y}F_{p,i}^s$ to denote the average value of the resulting control gain from $Y$ experiments, where $F_{p,i}^s$ is the control gain obtained at the $s$-th step from the $i$-th experiment.
Correspondingly, the average deviation of the approximation error at the $s$-th iteration step obtained from $Y$ experiments is denoted by  $\text{E}_p^s=\frac{1}{Y}\sum_{i=1}^Y \|F_{p,i}^s-F^*\|$.
In Fig.~\ref{fig:alg3-sigma}, we show the value of $F_p^s$ at each experiment as well as the resulting average value $\check{F}_p^s$, when the variance of the system noises is set as $\sigma_w^2=\textbf{0}_2$, $\sigma_w^2=0.1\textbf{I}_2$, $\sigma_w^2=0.5\textbf{I}_2$, and $\sigma_w^2=\textbf{I}_2$, respectively.
By observation, we know that in these four cases, $F_{p}$ is in a small neighborhood of the optimal feedback gain after about $3$ iterations, and the range of the neighborhood is increased as the increase of the variance of the system noise. 
This implies that when the variance of the system noise is large, Algorithm 5 may not converge under a small convergence precision.
In spite of this, we can also see from Fig.~\ref{fig:alg3-sigma} that the average control gain $\check{F}^s$ is close to the optimal one.
Let $E_p^{15}=\frac{1}{Y}\sum_{i=1}^Y\|F_{p,i}^{15}-F^*\|$, which is the average derivation of the control gain obtained at the $15$-th iteration step of the $i$-th experiment from the optimal control gain $F^*$.
Fig.~\ref{fig:alg5-sigma-ave-trend} shows the values of $E_p^{15}$ when $\alpha$ is increased from $0$ to $1$.
Roughly speaking, the resulting approximation errors are increased as the increase of the variance of the system noise.

\begin{figure}
	\centering
	\vspace{-0cm} 
	\subfigtopskip=0pt 
	\subfigbottomskip=0pt 
	\subfigcapskip=-2pt 
	\subfigure[\hspace{-0.5cm}]
	{
		\includegraphics[width=3.8cm]{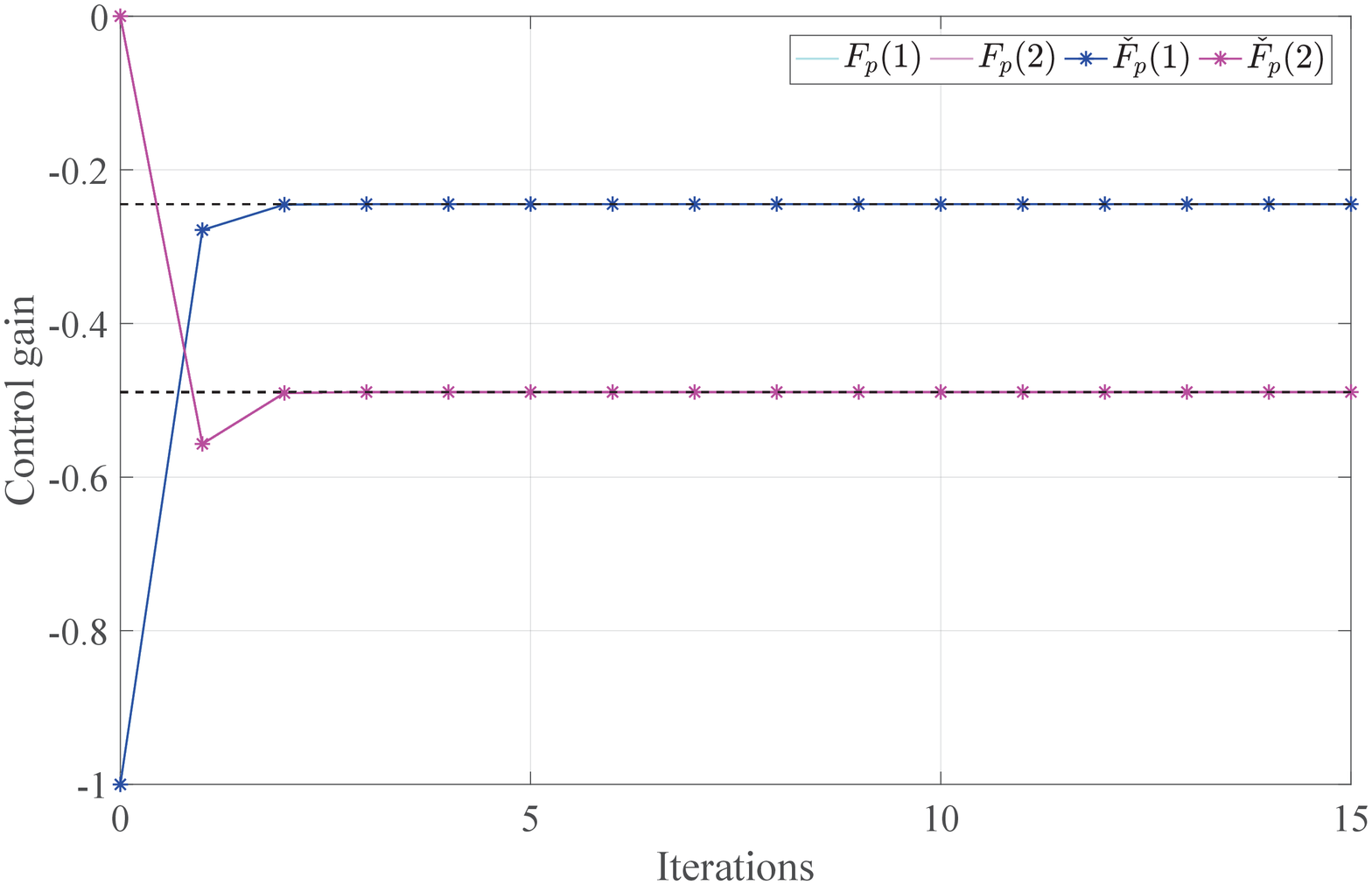}
		\label{fig:alg5-sigma0}
	}
	\subfigure[\hspace{-0.5cm}]
	{
		\includegraphics[width=3.8cm]{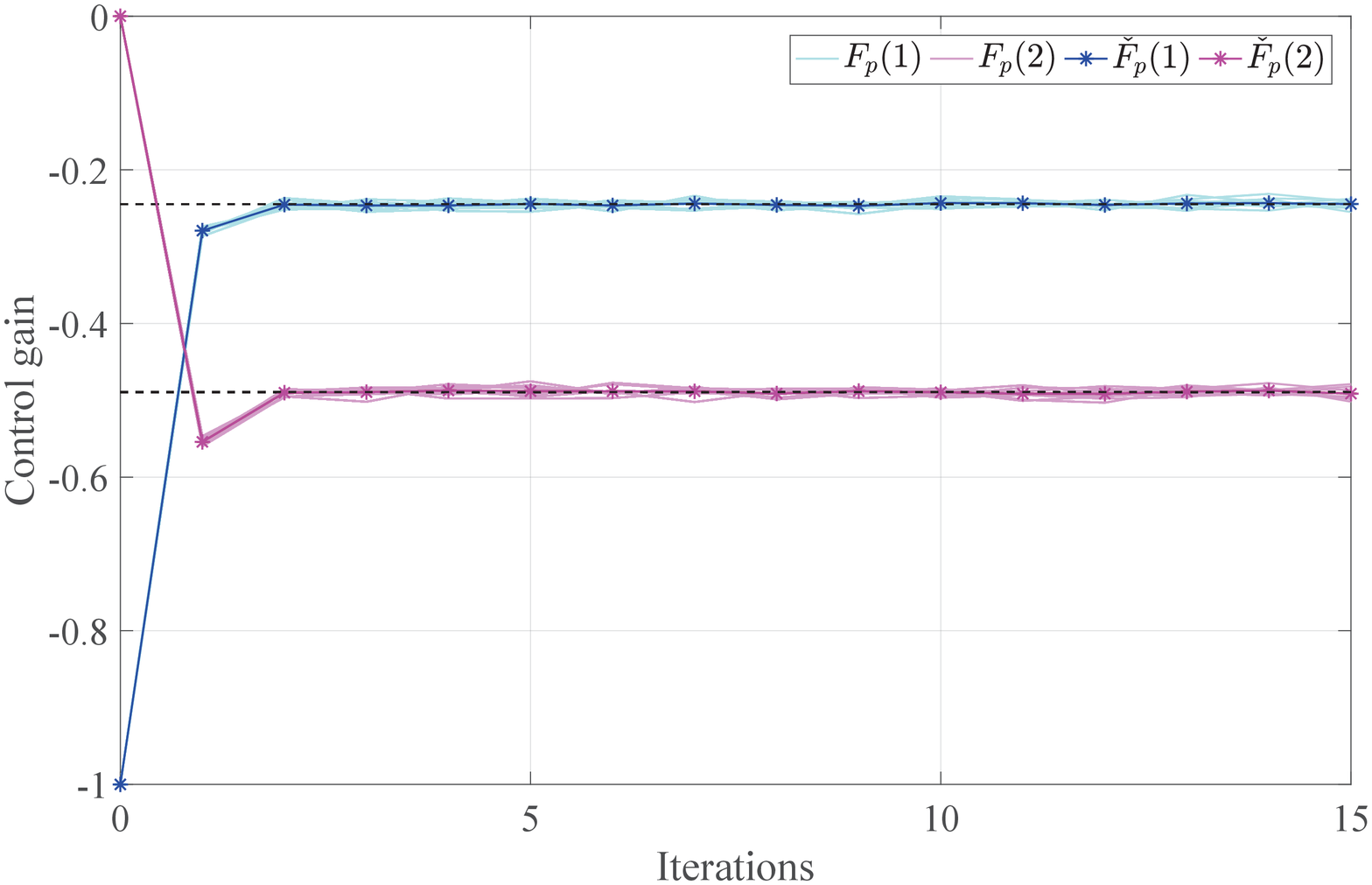}
		\label{fig:alg5-sigma01}
	}
	\subfigure[\hspace{-0.5cm}]
	{
		\includegraphics[width=3.8cm]{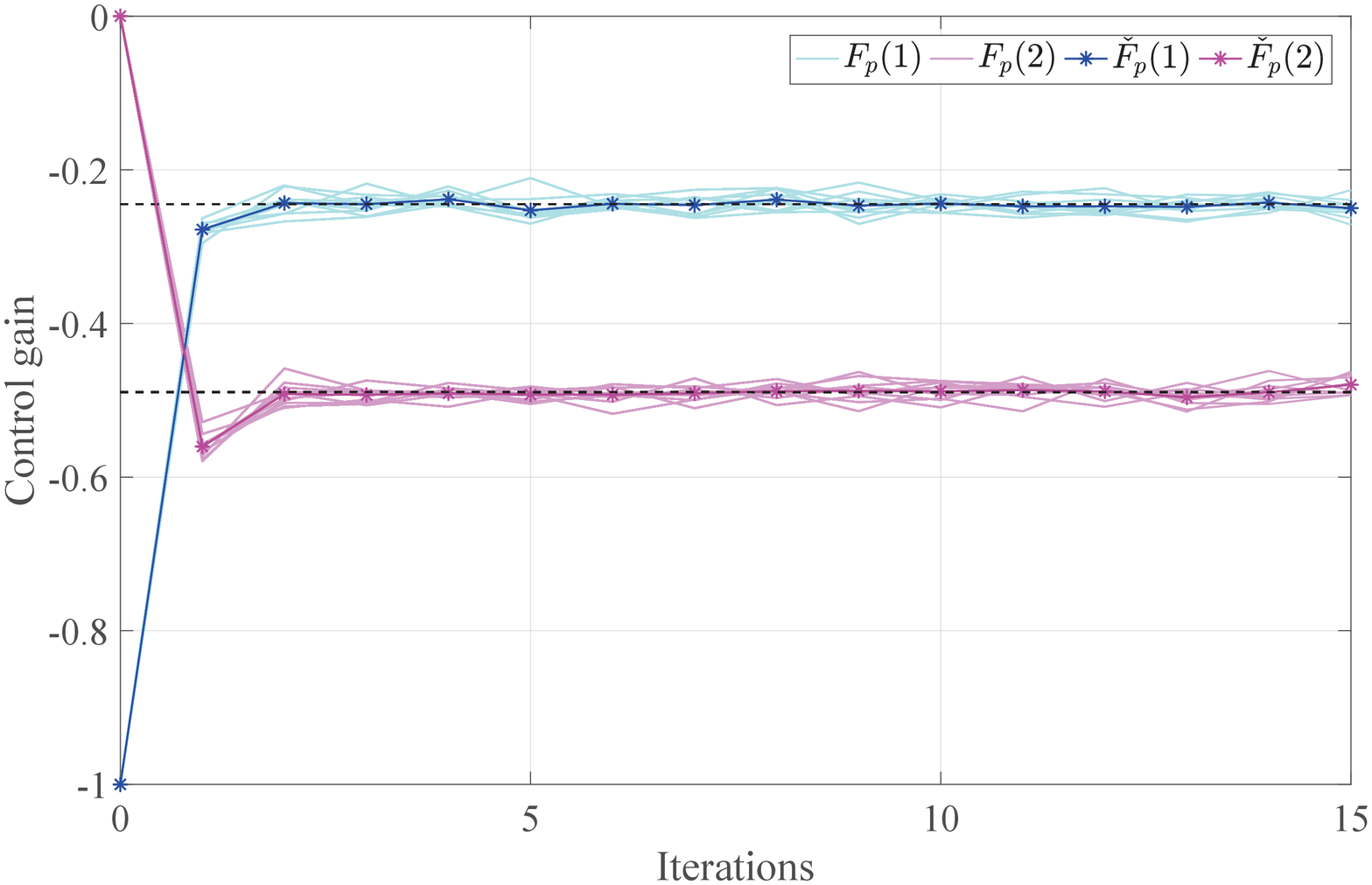}
		\label{fig:alg5-sigma05}
	}
	\subfigure[\hspace{-0.5cm}]
	{
		\includegraphics[width=3.8cm]{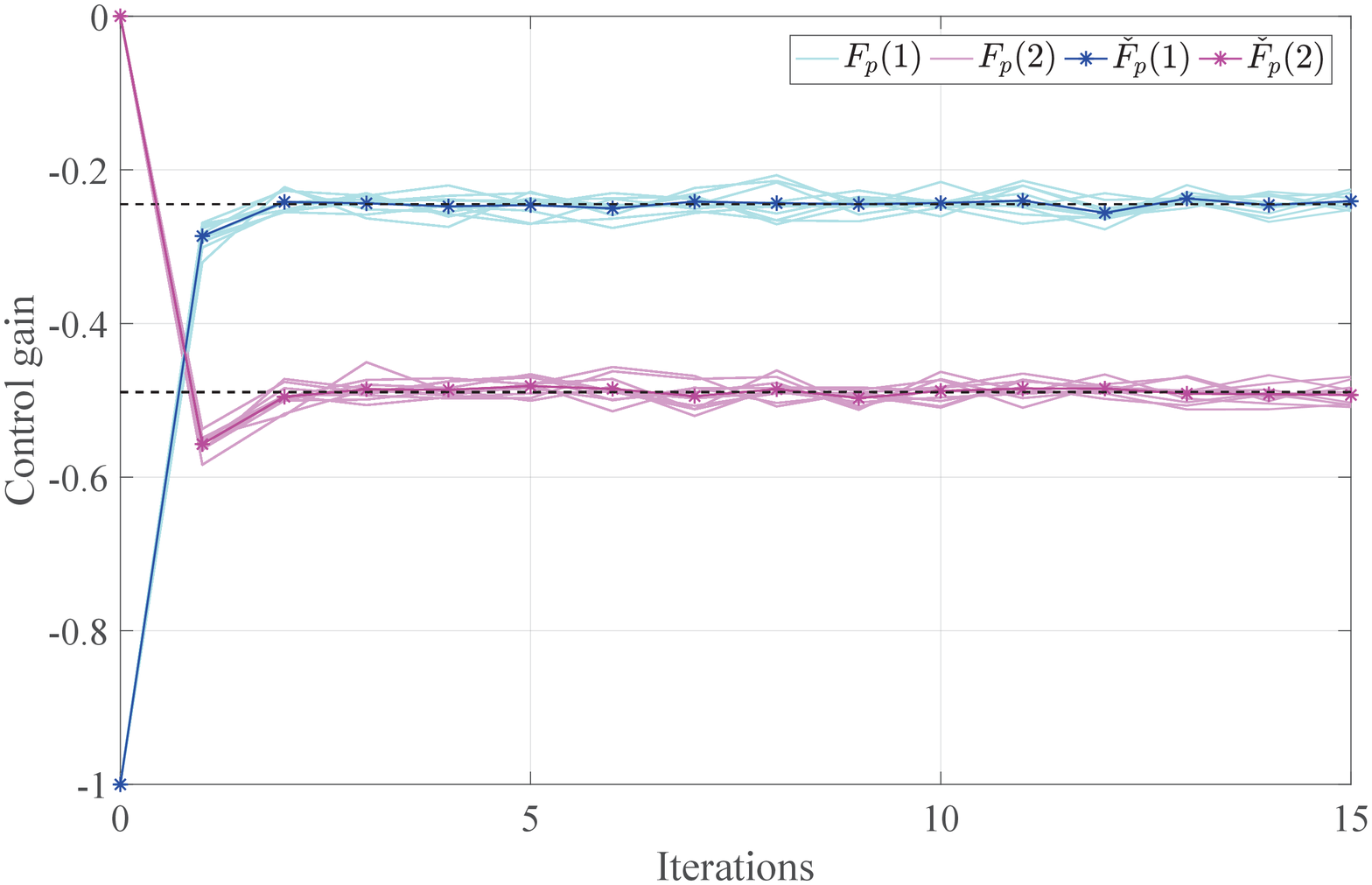}
		\label{fig:alg5-sigma1}
	}
	\caption{Values of $F_p^{s}$ at each iteration step. The black dotted lines are baselines for two components of the optimal control gain $F^*$. (a) $\sigma_w^2=\textbf{0}_2$. (b) $\sigma_w^2=0.1\textbf{I}_2$. (c) $\sigma_w^2=0.5\textbf{I}_2$. (d) $\sigma_w^2=\textbf{I}_2$.}
	\label{fig:alg5-sigma}
\end{figure}

\section{Conclusion}\label{concl}
\vspace{-2mm}
In this paper, we have revisited the stochastic LQR problem, and have developed a MF-OPPI algorithm and a MB-PD algorithm from the perspective of RL and primal-dual optimization, respectively.
Both algorithms can be implemented without using the information of system matrices, such that one can circumvent model identification, which may cause undesirable behaviors due to the propagation and accumulation of modeling errors.
Furthermore, we have shown that the dual and primal update steps in the MB-PD algorithm can be interpreted as the policy evaluation and policy improvement steps in the classical PI algorithm.
This advances our understanding of common RL algorithms and might contribute to develop various RL algorithms based on optimization formulations.


\appendix
\section{Proof of Theorem~\ref{thm-1}}\label{proof_thm1}
\vspace{-2mm}
From the proof of Proposition~\ref{equiv-primal-1}, we know that the constraints $S\succ 0$ and $F\in\mathcal{F}$ are equivalent.
Based on this, we introduce the following auxiliary problem for further analysis:
\vspace{-2mm}
\begin{auxiliary-pro}\label{primal-3}
	Solve \vspace{-2mm}
	\begin{align}
	\hat{J}_p\triangleq &\inf_{S\in\mathbb{S}^{n+m},F\in\mathbb{R}^{m\times n}} \tr (\Lambda S)\notag\\
	\st \;\; & F\in\mathcal{F},\label{primal-1-nonconvex-cons}\\
	& \gamma A_FSA_F^{\prime}+\Gamma+\frac{\gamma}{1-\gamma}\bar{F}\sigma_w^2\bar{F}^{\prime}=S.\label{primal-1-nonconvex-cons1}
	\end{align}
\end{auxiliary-pro}

\vspace{-6mm} This is a non-convex optimization problem since the set $\mathcal{F}$ is not convex and the equality constraint \eqref{primal-1-nonconvex-cons1} is a quadratic function of $F$.
Note that we can easily deduce from the proof of Proposition~\ref{equiv-primal-1} that the optimal value of Auxiliary Problem~\ref{primal-3} is equivalent to that of Problem~\ref{LQG-problem2}, i.e., $\hat{J}_p=\hat{J}^*$.

\vspace{-4mm} The Lagrangian function of Auxiliary Problem~\ref{primal-3} is defined as \vspace{-2mm}
\begin{equation}
\begin{aligned}\label{Lagrangian-1}
&\hat{L}(P,F,S)\\
\triangleq&\tr(\Lambda S)+\tr\left[\Big(\gamma A_FSA_F^{\prime}+\Gamma + \frac{\gamma}{1-\gamma}\bar{F}\sigma_w^2\bar{F}^{\prime}-S\Big)P\right],
\end{aligned}
\end{equation}

\vspace{-3mm} \noindent where $P\in\mathbb{S}^{n+m}$ is the Lagrange multiplier. 
The corresponding Lagrangian dual function is defined as \vspace{-2mm}
\begin{align*}
\hat{d}(P)\triangleq \inf_{S\in\mathbb{S}^{n+m},F\in\mathcal{F}}\hat{L}(P,F,S).
\end{align*}

\vspace{-6mm} \noindent 
We take the constraint $F\in\mathcal{F}$ as the domain of the variable $F$ in the Lagrangian formulation, instead of directly writing it in the Lagrangian function, since  the constraint $F\in\mathcal{F}$ is not an explicit equality or inequality constraint. \vspace{-4mm}

Then, the dual problem of Auxiliary Problem~\ref{primal-3} is given as follows:\vspace{-2mm}
\begin{align*}
\hat{J}_d \triangleq \sup_{P\in\mathbb{S}^{n+m}} \hat{d}(P) =\sup_{P\in\mathbb{S}^{n+m}} \inf_{S\in\mathbb{S}_{+}^{n+m},F\in\mathcal{F}}\hat{L}(P,F,S),
\end{align*}

\vspace{-6mm} \noindent  where $\hat{L}(P,F,S)$ is as defined in \eqref{Lagrangian-1}.\vspace{-4mm}

Since $J_p=\hat{J}^*$ and $\hat{J}_p=\hat{J}^*$, it holds that $J_p=\hat{J}_p$.
By weak duality, we have $J_p\ge J_d$.
In view of this, we complete the proof of Theorem~\ref{thm-1} by the following two parts:\vspace{-3mm}
\begin{enumerate}
	\item Auxiliary Problem~\ref{primal-3} has strong duality, i.e., $\hat{J}_p=\hat{J}_d$ (Subsection~\ref{appenA-1});
	\item The optimal point of the dual problem of Primal Problem~\ref{primal-1} is not larger than that of the dual problem of Auxiliary Problem~\ref{primal-3}, i.e., $J_d\le\hat{J}_d$ (Subsection~\ref{appenA-2}).
\end{enumerate}\vspace{-2mm}

\subsection{Proof for $\hat{J}_p=\hat{J}_d$}\label{appenA-1}
\vspace{-2mm}
We first introduce another auxiliary optimization problem which can be shown to be equivalent to the modified stochastic LQR problem, i.e., Problem~\ref{LQG-problem2}.\vspace{-3mm}  
\begin{auxiliary-pro}\label{primal-2}
	Solve\vspace{-2mm}
	\begin{align}
	\tilde{J}_{p}\triangleq&\inf_{P\in\mathbb{S}^{n+m},F\in\mathbb{R}^{m\times n}} \tr\left(\Big(\Gamma+\frac{\gamma}{1-\gamma}\bar{F}\sigma_w^2\bar{F}^{\prime}\Big)P\right)\notag\\
	\st \;\; &F\in\mathcal{F},\label{primal-2-nonconvex}\\
	&\gamma A_F^{\prime}PA_F+\Lambda=P. \label{primal-2-Lya}
	\end{align}
\end{auxiliary-pro}  

\vspace{-6mm}\begin{proposition}\label{equiv-primal-2}
	The Auxiliary Problem~\ref{primal-2} has a unique optimal solution denoted by $(\tilde{P}_{p},\tilde{F}_{p})$, and it is equivalent to Problem~\ref{LQG-problem2} in the sense that $\tilde{J}_{p}=\hat{J}^*$ and $\tilde{F}_{p}=\hat{F}^*$.
\end{proposition}\vspace{-3mm}  
\begin{proof}
	By the equation \eqref{v} and the fact that $w_k\sim \mathcal{N}(0_n,\sigma_w^2)$ is the i.i.d. random noise for each $k\ge 0$, we can obtain that for any $F\in\mathcal{F}$, it holds that\vspace{-2mm}  
	\begin{align*}
	&\mathbb{E}\left[v_k(F,v_{(i)})^{\prime}\Lambda v_k(F,v_{(i)})\right]\\
	=&\tr\left[\mathbb{E}\left(v_{(i)}v_{(i)}^{\prime}\right)(A_F^k)^{\prime}\Lambda A_F^k\right]+\tr\left(\sum_{j=0}^{k-1}\bar{F}\sigma_w^2\bar{F}^{\prime}(A_F^j)^{\prime}\Lambda A_F^j\right).
	\end{align*}
	
	\vspace{-6mm} \noindent Therefore, the objective function of Problem~\ref{LQG-problem2} becomes\vspace{-2mm}  
	\begin{align*}
	\frac{1}{r}\sum_{i=1}^r \hat{J}(F,v_{(i)})=\tr\left(\Gamma P_1\right)+\frac{\gamma}{1-\gamma}\tr\left( \bar{F}\sigma_w^2 \bar{F}^{\prime} P_2 \right),
	\end{align*}
	
	\vspace{-6mm} \noindent with\vspace{-2mm}  
	\begin{align*}
	P_1&=\sum_{k=0}^{\infty} \gamma^k (A_F^k)^{\prime}\Lambda A_F^k,\\
	P_2&=\frac{1-\gamma}{\gamma}\sum_{k=0}^{\infty}\gamma^k\sum_{j=0}^{k-1}(A_F^j)^{\prime}\Lambda A_F^j.
	\end{align*}
	
	\vspace{-6mm} \noindent We can easily verify that both $P_1$ and $P_2$ satisfy the Lyapunov equation \eqref{primal-2-Lya}.
	By the first statement of Lemma~\ref{Lyapunov}, there exists a unique $P\in\mathbb{S}^n_{+}$ such that $\gamma A_F^{\prime}PA_F+\Lambda=P$, which implies that $P_1=P_2$.\vspace{-4mm}  
	
	Since the optimal solution $(\tilde{P}_p,\tilde{F}_p)$ of Auxiliary Problem~\ref{primal-2} is one feasible point of Problem~\ref{LQG-problem2}, it holds that $\tilde{J}_p\ge \hat{J}^*$.
	Furthermore, if $\tilde{F}_{p}=\hat{F}^*\in\mathcal{F}$, and $\tilde{P}_p$ is the unique solution to the constraint \eqref{primal-2-Lya}, then the resulting objective function of Auxiliary Problem~\ref{primal-2} is exactly $\hat{J}^*$.
	Therefore, we can conclude that $(\tilde{S}_p,\tilde{F}_p)$ is the solution of Auxiliary Problem~\ref{primal-2} and $\tilde{J}_p=\hat{J}^*$.
	The uniqueness of $(\tilde{S}_p,\tilde{F}_p)$ can be obtained by the uniqueness of $\hat{F}^*$.\vspace{-3mm}
\end{proof}

To show the strong duality of Auxiliary Problem~\ref{primal-3}, the following lemmas will be used.\vspace{-2mm}  
\begin{lemma}\label{primal-2-prop}
	The optimal point $(\tilde{P}_p,\tilde{F}_p)$ of Auxiliary Problem~\ref{primal-2} satisfies the following:\vspace{-2mm}
	\begin{enumerate}
		\item $\tilde{P}_{p}=P^*$;
		\item $\tilde{P}_{p}\succeq 0$, $\tilde{P}_{p,22}\succ 0$, and $\tilde{F}_{p}=-(\tilde{P}_{p,22})^{-1}(\tilde{P}_{p,12})^{\prime}\in\mathcal{F}$.
	\end{enumerate}
\end{lemma}\vspace{-2mm}
\begin{proof}
	Based on the definitions in \eqref{optimal-F} and \eqref{P}, direct calculation yields that $\left[\begin{array}{c}
	I_n\\
	F^*
	\end{array}\right]^{\prime}P^*\left[\begin{array}{c}
	I_n\\
	F^*
	\end{array}\right]=X^*$.
	Then, from \eqref{P}, it follows that $P^*=\gamma \left[\begin{array}{cc}
	A & B
	\end{array}\right]^{\prime}X^*\left[\begin{array}{cc}
	A & B
	\end{array}\right]+\Lambda=\gamma \left[\begin{array}{cc}
	A & B
	\end{array}\right]^{\prime}\left[\begin{array}{c}
	I_n\\
	F^*
	\end{array}\right]^{\prime}P^*\left[\begin{array}{c}
	I_n\\
	F^*
	\end{array}\right]\left[\begin{array}{cc}
	A & B
	\end{array}\right]+\Lambda=\gamma A_{F*}^{\prime}P^*A_{F*}+\Lambda$.
	Since $F^*\in\mathcal{F}$, i.e., $\rho(A_{F^*})<1$, there exists a unique $P^*$ satisfying the above equation.\vspace{-4mm}
	
	In addition, we have $\tilde{F}_{p}=F^*=\hat{F}^*$ by Proposition~\ref{equiv-F} and Proposition~\ref{equiv-primal-1}.
	Therefore, when $F=\tilde{F}_{p}$, the equality constraint \eqref{primal-2-Lya} has a unique solution $\tilde{P}_{p}=P^*$.\vspace{-4mm}
	
	The second statement can be directly shown via the definitions of $P^*$ and $F^*$ given in \eqref{P} and \eqref{optimal-F}, respectively.
\end{proof}\vspace{-3mm}

\begin{lemma}[{\color{black}\cite[Lemma~4]{lee2018primal}}]\label{lemma-for-thm1}
	If $P\succeq 0$ and $P_{22}\succeq 0$, then for $\forall F\in\mathbb{R}^{m\times n}$, it holds that \vspace{-2mm}
	\begin{align*}
	\bar{F}^{\prime}P\bar{F}\succeq P_{11}-P_{12}P_{22}^{-1}P_{12}^{\prime}=\left[
	\begin{array}{c}
	I_n\\
	P_{22}^{-1}P_{12}^{\prime}
	\end{array}\right]^{\prime}P\left[
	\begin{array}{c}
	I_n\\
	P_{22}^{-1}P_{12}^{\prime}
	\end{array}\right],
	\end{align*}
	
	\vspace{-6mm} \noindent and the equality holds if and only if $F=-P_{22}^{-1}P_{12}^{\prime}$.
\end{lemma}\vspace{-3mm}

Now, we are ready to show the strong duality of Auxiliary Problem ~\ref{primal-3}.\vspace{-4mm}
\begin{lemma}\label{Jp=Jd}
	The dual gap for Auxiliary Problem ~\ref{primal-3} is zero, that is, $\hat{J}_p=\hat{J}_d$.
\end{lemma}\vspace{-3mm}
\begin{proof}
	For a given matrix $P\in\mathbb{S}^{n+m}$, the Lagrangian dual function is \vspace{-2mm}
	\begin{align*}
	\hat{d}(P)&=\inf_{S\in\mathbb{S}_{+}^{n+m},F\in\mathcal{F}}\hat{L}(P,F,S)\\
	&=\left\{
	\begin{aligned}
	&\inf_{F\in\mathcal{F}} \tr\left( \Gamma P \right)+\frac{\gamma}{1-\gamma}\tr\left( \sigma_w^2\bar{F}^{\prime} P\bar{F} \right), \; \text{if} \ P\in\hat{\mathcal{P}}\\
	& -\infty, \; \text{otherwise}
	\end{aligned}
	\right.
	\end{align*}
	
	\vspace{-6mm} \noindent where $\hat{\mathcal{P}}=\{P\in\mathbb{S}^{n+m}: \gamma A_F^{\prime}PA_F-P+\Lambda\succeq 0, \forall F\in\mathcal{F} \}$. \vspace{-4mm}
	
	Next, we show that the optimal solution $\tilde{P}_{p}$ of Auxiliary Problem~\ref{primal-2} is an element of $\hat{\mathcal{P}}$, and thus the set $\hat{\mathcal{P}}$ is nonempty.
	By Lemma~\ref{primal-2-prop}, $\gamma A_{\tilde{F}_{p}}^{\prime}\tilde{P}_{p}A_{\tilde{F}_{p}}+\Lambda=\tilde{P}_{p}$, where $\tilde{F}_{p}=-(\tilde{P}_{p,22})^{-1}(\tilde{P}_{p,12})^{\prime}$.
	Then, by Lemma~\ref{lemma-for-thm1}, it holds that $\gamma A_{F}^{\prime}\tilde{P}_{p}A_{F}+\Lambda\succeq \gamma A_{\tilde{F}_{p}}^{\prime}\tilde{P}_{p}A_{\tilde{F}_{p}}+\Lambda=\tilde{P}_{p}$ for all $F\in\mathbb{R}^{m\times n}$.\vspace{-4mm}
	
	Therefore, the dual problem is equivalent to \vspace{-2mm}
	\begin{equation}\label{appen1}
	\begin{aligned}
	\hat{J}_d=&\sup_{P\in\mathbb{S}^{n+m}} \hat{d}(P)\\
	=&\sup_{P\in\hat{\mathcal{P}}} \inf_{F\in\mathcal{F}} \tr(\Gamma P)+\frac{\gamma}{1-\gamma}\tr(\sigma_w^2\bar{F}^{\prime}P\bar{F}).
	\end{aligned}
	\end{equation}
	
	\vspace{-6mm} \noindent  For $\tilde{P}_{p}\in\hat{\mathcal{P}}$, we have\vspace{-2mm}
	\begin{equation}
	\begin{aligned}\label{thm1-eq1}
	\hat{d}(\tilde{P}_{p})=\inf_{F\in\mathcal{F}} \tr(\Gamma \tilde{P}_{p})+\frac{\gamma}{1-\gamma}\tr(\sigma_w^2\bar{F}^{\prime}\tilde{P}_{p}\bar{F}).
	\end{aligned}
	\end{equation}
	
	\vspace{-6mm}\noindent Obviously, $\hat{d}(\tilde{P}_{p})\le \hat{J}_d$.
	Since $\tilde{P}_{p}\succeq 0$ is fixed and the objective function in \eqref{thm1-eq1} is a quadratic function of $F$, the infimum in \eqref{thm1-eq1} is reached at $F=-(\tilde{P}_{p,22})^{-1}(\tilde{P}_{p,12})^{\prime}=\tilde{P}_{p}\in\mathcal{F}$.
	Therefore, $\tilde{J}_{p}=\hat{d}(\tilde{P}_{p})$, which implies that $\tilde{J}_{p}\le \hat{J}_d$.
	By the weak duality, there holds that $\hat{J}_d\le \hat{J}_p$.
	Since $\hat{J}_p=\hat{J}^*=\tilde{J}_{p}$, we have $\hat{J}_p=\hat{J}_d$.\vspace{-2mm}
\end{proof}

\subsection{Proof for $J_d\le \hat{J}_d$}\label{appenA-2}
\vspace{-2mm}

\begin{lemma}
	There holds $J_d\le\hat{J}_d$.
\end{lemma}\vspace{-3mm}
\begin{proof}
	For any given $P\in\mathbb{S}^{n+m}$, $P_0\in\mathbb{S}_{+}^{n+m}$, the Lagrangian dual function $d(P,P_0)$ is \vspace{-2mm}
	\begin{align*}
	d(P,P_0)=&\inf_{S\in\mathbb{S}_{+}^{n+m},F\in\mathbb{R}^{m\times n}}L (P,P_0,F,S)\\
	=&\left\{
	\begin{aligned}
	&\inf_{F\in\mathbb{R}^{m\times n}} \tr\left( \Gamma P \right)+\frac{\gamma}{1-\gamma}\tr\left( \sigma_w^2\bar{F}^{\prime} P\bar{F} \right), \; \text{if} \ P\in\hat{\mathcal{P}}\\
	& -\infty, \; \text{otherwise}
	\end{aligned}
	\right.
	\end{align*}
	
	\vspace{-6mm} \noindent where $\mathcal{P}=\{P\in\mathbb{S}_{+}^{m+n}:\gamma A_F^{\prime}PA_F-P+\Lambda-P_0\succeq 0, \forall F\in\mathcal{F} \}$. \vspace{-4mm}
	
	Therefore, the dual problem corresponding to Primal Problem~\ref{primal-1} is equivalent to \vspace{-2mm}
	\begin{equation}
	\begin{aligned}\label{appen2}
	J_d=&\sup_{P\in\mathbb{S}_{+}^{m+n}, P_0\in\mathbb{S}_{+}^{m+n}} d(P,P_0)\\
	=&\sup_{P\in\mathcal{P}, P_0\in\mathbb{S}_{+}^{m+n}} \inf_{F\in\mathbb{R}^{m\times n}} \tr\left( \Gamma P \right)+\frac{\gamma}{1-\gamma}\tr\left( \sigma_w^2\bar{F}^{\prime} P\bar{F} \right).
	\end{aligned}
	\end{equation}
	
	\vspace{-6mm} We can observe from \eqref{appen2} that $P_0$ affects the dual optimal value $J_d$ by adjusting the size of the set $\mathcal{P}$.
	To obtain the supremum of the Lagrangian dual function, the constraint $P\in\mathcal{P}$ should be relaxed as much as possible.
	We can easily observe from the definition of the set $\mathcal{P}$ that its maximum is obtained when $P_0=\textbf{0}_{m+n}$. 
	Furthermore, if $P_0=\textbf{0}_{m+n}$, then it holds that $\mathcal{P}=\{P\in\mathbb{S}_{+}^{m+n}:\gamma A_F^{\prime}PA_F-P+\Lambda\succeq 0, \forall F\in\mathcal{F}\}=\hat{\mathcal{P}}$.
	As a result, it yields that \vspace{-2mm}
	\begin{align*}
	J_d=\sup_{P\in\hat{\mathcal{P}}} \inf_{F\in\mathbb{R}^{m\times n}} \tr\left( \Gamma P \right)+\frac{\gamma}{1-\gamma}\tr\left( \sigma_w^2\bar{F}^{\prime} P\bar{F} \right).
	\end{align*}
	
	\vspace{-6mm} Since $\mathcal{F} \subseteq \mathbb{R}^{m\times n}$, it holds that $\inf_{F\in\mathbb{R}^{m\times n}} \tr\left( \Gamma P \right)+\frac{\gamma}{1-\gamma}\tr\left( \sigma_w^2\bar{F}^{\prime} P\bar{F} \right) \le \inf_{F\in\mathcal{F}} \tr\left( \Gamma P \right)+\frac{\gamma}{1-\gamma}\tr\left( \sigma_w^2\bar{F}^{\prime} P\bar{F} \right)$, which implies that $J_d\le\hat{J}_d$.
\end{proof}

\begin{thebibliography}{10}
	\expandafter\ifx\csname url\endcsname\relax
	\def\url#1{\texttt{#1}}\fi
	\expandafter\ifx\csname urlprefix\endcsname\relax\def\urlprefix{URL }\fi
	\expandafter\ifx\csname href\endcsname\relax
	\def\href#1#2{#2} \def\path#1{#1}\fi
	
	\bibitem{sutton1998reinforcement}
	R.~S. Sutton, A.~G. Barto. (1998). Reinforcement Learning: An introduction, MIT Press,
	Cambridge, MA, USA.
	
	\bibitem{bertsekas2019reinforcement}
	D.~P. Bertsekas. (2019). Reinforcement {Learning} and {Optimal} {Control}, Athena
	Scientific, Nashua, USA.
	
	\bibitem{Arulkumaran2017deep}
	K.~Arulkumaran, M.~P. Deisenroth, M.~Brundage, A.~A. Bharath. (2017). Deep
	reinforcement learning: A brief survey, IEEE Signal Processing Magazine,
	34~(6), 26--38.
	
	\bibitem{Recht2019atour}
	B.~Recht. (2019). A tour of reinforcement learning: The view from continuous control,
	Annual Review of Control, Robotics, and Autonomous Systems, 2, 253--279.
	
	\bibitem{qin2020multiplayer}
	M.~Li, J.~Qin, N.~M. Freris, D.~W.~C. Ho, Multiplayer {Stackelberg-Nash} game
	for nonlinear system via value iteration-based integral reinforcement
	learning, IEEE Transactions on Neural Networks and Learning Systems, DOI:
	10.1109/TNNLS.2020.3042331.
	
	\bibitem{li2019distributed}
	F.~Li, J.~Qin, W.~X. Zheng. (2019). Distributed {Q}-learning-based online optimization
	algorithm for unit commitment and dispatch in smart grid, IEEE Transactions
	on Cybernetics, 50~(9), 4146--4156.
	
	\bibitem{wang2020mobile}
	B.~Wang, Z.~Liu, Q.~Li, A.~Prorok. (2020). Mobile robot path planning in dynamic
	environments through globally guided reinforcement learning, IEEE Robotics
	and Automation Letters, 5~(4), 6932--6939.
	
	\bibitem{Haydari2020deep}
	A.~Haydari, Y.~Yilmaz, Deep reinforcement learning for intelligent
	transportation systems: A survey, IEEE Transactions on Intelligent
	Transportation Systems, DOI: 10.1109/TITS.2020.3008612.
	
	\bibitem{Leong2020deep}
	A.~S. Leong, A.~Ramaswamy, D.~E. Quevedo, H.~Karl, L.~Shi, Deep reinforcement
	learning for wireless sensor scheduling in cyber–physical systems,
	Automatica, DOI: https://doi.org/10.1016/j.automatica.2019.108759.
	
	\bibitem{naik2019discounted}
	A.~Naik, R.~Shariff, N.~Yasui, R.~S. Sutton, Discounted reinforcement learning
	is not an optimization problem, arXiv preprint arXiv:1910.02140.
	
	\bibitem{zhang2019text}
	R.~Zhang, T.~Yu, Y.~Shen, H.~Jin, C.~Chen. (2019). Text-based interactive
	recommendation via constraint-augmented reinforcement learning, in: Advances
	in Neural Information Processing Systems, Vancouver, Canada, pp.
	15214--15224.
	
	\bibitem{qin2014sparse}
	Z.~Qin, W.~Li, F.~Janoos. (2014). Sparse reinforcement learning via convex
	optimization, in: International Conference on Machine Learning, Beijing,
	China, pp. 424--432.
	
	\bibitem{Vieillard2019on}
	N.~Vieillard, O.~Pietquin, M.~Geist, On connections between constrained
	optimization and reinforcement learning, arXiv preprint arXiv:1910.08476.
	
	\bibitem{schluter2020event}
	H.~Schl\"uter, F.~Solowjow, S.~Trimpe, Event-triggered learning for linear
	quadratic control, IEEE Transactions on Automatic Control, DOI:
	10.1109/TAC.2020.3030877.
	
	\bibitem{Fazel2018global}
	M.~Fazel, R.~Ge, S.~M. Kakade, M.~Mesbahi. (2018). Global convergence of policy
	gradient methods for the linear quadratic regulator, in: International
	Conference on Machine Learning, Stockholm, Sweden, pp. 1467--1476.
	
	\bibitem{Malik2019derivative}
	D.~Malik, A.~Pananjady, K.~Bhatia, K.~Khamaru, B.~L. Peter, W.~J. Martin. (2019).
	Derivative-free methods for policy optimization: {Guarantees} for linear
	quadratic systems, in: The 22nd International Conference on Artificial
	Intelligence and Statistics, Stockholm, Sweden, pp. 2916--2925.
	
	\bibitem{li2020distributed}
	Y.~Li, Y.~Tang, R.~Zhang, N.~Li, Distributed reinforcement learning for
	decentralized linear quadratic control: A derivative-free policy optimization
	approach, DOI: arXiv preprint arXiv:1912.09135.
	
	\bibitem{karl2019finite}
	K.~Karl, S.~Tu. (2019). Finite-time analysis of approximate policy iteration for the
	linear quadratic regulator, in: International Conference on Machine Learning,
	Vancouver, Canada, pp. 8514--8524.
	
	\bibitem{Yang2019provably}
	Z.~Yang, Y.~Chen, M.~Hong, Z.~Wang. (2019). Provably global convergence of
	actor-critic: {A} case for linear quadratic regulator with ergodic cost, in:
	Advances in Neural Information Processing Systems, Vancouver, Canada, pp. 8353--8365.
	
	\bibitem{mitze2020adynamic}
	R.~Mitze, M.~M\"onnigmann, A dynamic programming approach to solving
	constrained linear–quadratic optimal control problems, Automatica, DOI:
	https://doi.org/10.1016/j.automatica.2020.109132.
	
	\bibitem{lee2018primal}
	D.~{Lee}, J.~{Hu}. (2018). Primal-dual {Q}-learning framework for {LQR} design, IEEE
	Transactions on Automatic Control, 64~(9), 3756--3763.
	
	\bibitem{david2001stochastic}
	D.~D. Yao, S.~{Zhang}, X.~Y. {Zhou}. (2001). Stochastic linear-quadratic control via
	semidefinite programming, SIAM Journal on Control and Optimization, 40~(3), 801--823.
	
	\bibitem{jiang2012computational}
	Y.~Jiang, Z.-P. Jiang. (2012). Computational adaptive optimal control for
	continuous-time linear systems with completely unknown dynamics, Automatica,
	48~(10), 2699--2704.
	
	\bibitem{kiumarsi2017h}
	B.~Kiumarsi, F.~L. Lewis, Z.-P. Jiang. (2017). ${H}_{\infty}$ control of linear
	discrete-time systems: Off-policy reinforcement learning, Automatica, 78,
	 144--152.
	
	\bibitem{qin2018optimal}
	J.~Qin, M.~Li, Y.~Shi, Q.~Ma, W.~X. Zheng. (2019). Optimal synchronization control of
	multiagent systems with input saturation via off-policy reinforcement
	learning, IEEE Transactions on Neural Networks and Learning Systems, 30~(1), 85--96.
	
	\bibitem{xiong2020model}
	J.~{Lai}, J.~{Xiong}, Z.~{Shu}, Model-free optimal control of discrete-time
	systems with additive and multiplicative noises, arXiv preprint
	arXiv:2008.08734.
	
	\bibitem{vrabie2013optimal}
	D.~Vrabie, K.~G. Vamvoudakis, F.~L. Lewis. (2013). Optimal {A}daptive {C}ontrol and
	{D}ifferential {G}ames by {R}einforcement {L}earning {P}rinciples, IET Press,
	Stevenage, U.K.
	
	\bibitem{hewer1971an}
	G.~{Hewer}. (1971). An iterative technique for the computation of the steady state
	gains for the discrete optimal regulator, IEEE Transactions on Automatic
	Control, 16~(4), 382--384.
	
	\bibitem{graupe1980acom}
	D.~Graupe, V.~K. Jain, J.~Salahi. (1980). A comparative analysis of various
	least-squares identification algorithms, Automatica, 16~(6), 663--681.
	
	\bibitem{vrabie2009adaptive}
	D.~Vrabie, O.~Pastravanu, M.~Abu-Khalaf, F.~L. Lewis. (2009). Adaptive optimal control for continuous-time linear systems based on policy iteration, Automatica, 45~(2), 477--484.
	
	\bibitem{gu2012discrete}
	G.~Gu. (2012). Discrete-{T}ime {L}inear {S}ystems: {T}heory and {D}esign with
	{A}pplications, Springer-Verlag, London, U.K.
	
	\bibitem{boyd2004convex}
	S.~{Boyd}, B.~{Vandenberghe}. (2004). {Convex Optimization}, Cambridge University Press, Cambridge, U.K.
	
\end{thebibliography}

\end{document}